\documentclass[11pt,a4paper]{article}
\usepackage{jheppub,amsmath,amssymb,slashed,url,bm}
\usepackage{graphicx}
\usepackage{epstopdf}

\newcommand{\e}{\begin{eqnarray}}
\newcommand{\ee}{\end{eqnarray}}

\newcommand{\CN}{{\cal N}}

\newcommand{\DA}{{\dot A}}
\newcommand{\DB}{{\dot B}}
\newcommand{\DC}{{\dot C}}

\def\a{\alpha}
\def\b{\beta}
\def\d{\delta}
\newcommand{\ep}{\epsilon}
\newcommand{\g}{\gamma}

\newcommand{\s}{\sigma}

\newcommand{\vp}{\varepsilon}

\font\teneurm=eurm10 \font\seveneurm=eurm7  \font\fiveeurm=eurm5
\newfam\eurmfam
\textfont\eurmfam=\teneurm \scriptfont\eurmfam=\seveneurm
\scriptscriptfont\eurmfam=\fiveeurm

\font\teneusm=eusm10 \font\seveneusm=eusm7 \font\fiveeusm=eusm5
\newfam\eusmfam
\textfont\eusmfam=\teneusm \scriptfont\eusmfam=\seveneusm
\scriptscriptfont\eusmfam=\fiveeusm

\font\tencmmib=cmmib10 \skewchar\tencmmib='177
\font\sevencmmib=cmmib7 \skewchar\sevencmmib='177
\font\fivecmmib=cmmib5 \skewchar\fivecmmib='177
\newfam\cmmibfam
\textfont\cmmibfam=\tencmmib \scriptfont\cmmibfam=\sevencmmib
\scriptscriptfont\cmmibfam=\fivecmmib

\title{A
6D nonabelian $(1,0)$ Theory
}

 \author{Fa-Min Chen }
\affiliation{Department of Physics, Beijing Jiaotong University, Beijing 100044, China}
\abstract{We construct a 6D nonabelian ${\cal N}=(1,0)$ theory by coupling an ${\cal N}=(1, 0)$ tensor multiplet to an ${\cal N}=(1, 0)$ hypermultiplet.
While the  ${\cal N}=(1, 0)$ tensor multiplet is in the adjoint representation of the gauge group, the hypermultiplet can be in the fundamental representation or any other representation. If the hypermultiplet is also in the adjoint representation of the gauge group, the supersymmetry is enhanced to ${\cal N}=(2, 0)$, and the theory is identical to the $(2,0)$ theory of Lambert and Papageorgakis (LP). Upon dimension reduction, the $(1, 0)$ theory can be reduced to a general ${\cal N}=1$ supersymmetric Yang-Mills theory in 5D. We discuss briefly the possible applications of the theories to multi M5-branes.}
\begin{document} \maketitle

\section{Introduction and Summary}\label{secintro}
M2-branes and M5-branes are  two types of fundamental objects in M-theory. According to the gauge/gravity correspondence, they admit dual gauge descriptions \cite{maldacena97}.
The gauge theories of multi M2-branes have been constructed successfully: They are the 3D $\CN=8$ BLG theory with gauge group $SO(4)$ \cite{G09, BL08}, the $\CN=6$ ABJM theory with gauge group $U(N)\times U(N)$ \cite{ABJM}, and the other extended superconformal Chern-Simons matter theories with variety gauge groups. However, it seems more difficult to construct the gauge theory of multi M5-branes. One particular reason is that it is difficult to construct an action: The theory  contains a self-dual three-form field strength $H_{\mu\nu\rho}=\frac{1}{3!}\vp_{\mu\nu\rho\s\lambda\tau}H^{\s\lambda\tau}$, implying that the kinetic term $H_{\mu\nu\rho}H^{\mu\nu\rho}$ vanishies.

Fortunately, it is possible to construct the equations of motions and the laws of supersymmetry transformations of 6D $(r,0)$ theories. Here $r=1,2$.
Using a three-algebra approach, Lambert and Papageorgakis (LP) was able to derive a nonabelian $(2, 0)$ tensor multiplet theory \cite{LP}, which may be a candidate of the gauge description of multiple M5-branes (For reviews on gauge theories of M-branes, see \cite{bagger} and \cite{Lambert}). More recently, using the Nambu three-algebra, Lambert and Sacco (LS) have constructed a more general $(2,0)$ theory by introducing an additional non-dynamical abelian three-form into the LP theory \cite{LS}. Remarkably, upon a dimension reduction, the LS theory is reduced to the 3D $\CN=8$ BLG theory, describing two  M2-branes in $\mathbf{C}^4/\mathbf{Z}_2$. Thus the $(2,0)$ LS theory may be a dual gauge theory for  two M5-branes or two M2-branes. The LS theory has been investigated in Ref. \cite{KLO}, and an intereting solution was found in \cite{KLO}.

In this paper, we generalize the $(2,0)$ LP theory in another direction. We construct a 6D nonabelian $\CN=(1,0)$ theory  by coupling a ``minimal" $\CN=(1, 0)$ tensor multiplet to an $\CN=(1, 0)$ hypermultiplet. The ``minimal" $(1,0)$ tensor multiplet, constructed in our previous work \cite{Chen10}, is in the adjoint representation of the gauge group, but the $(1,0)$ hypermultiplet, can be in the fundamental representation or any other representation. The field content of theory is the same as that of the LP theory, but the R-symmetry is only $SU(2)$.  If the $(1,0)$ hypermultiplet also takes value in the adjoint representation, then the $SU(2)$ R-symmetry can be promoted to $SO(5)$, and the supersymmetry gets enhanced to $(2,0)$, and our theory becomes identical to the $(2,0)$ LP theory. However, if the hypermultiplet is \emph{not} in the adjoint representation of the gauge group, our theory is a real $(1,0)$ theory. In fact, if the tensor multiplet and hypermultiplet are in different representations, it is impossible to promote the $SU(2)$ R-symmetry to $SO(5)$, meaning that one cannot enhance the $(1,0)$ supersymmetry to $(2,0)$\footnote{In our previous work \cite{Chen10}, only the ``minimal" $(1,0)$ tensor multiplet theory is a genuine  $(1,0)$ theory (see Section 2 of \cite{Chen10}). After coupling to the hypermultiplet, which is also in the adjoint representation of the gauge group, the resulted $(1,0)$ tensor multiplet theory in Ref. \cite{Chen10} is \emph{not} a real $(1,0)$ theory with $SU(2)$ R-symmetry. A careful analysis can show that the parameter ``$b$" in Ref. \cite{Chen10} can be absorbed into the re-definition of the fields, and the theory turns out to be the $(2,0)$  LP theory; In other words, it is actually a re-derivation of $(2,0)$  LP theory using a different approach.}.

Following the method of \cite{LP}, we show that this $(1,0)$ theory can be reduced to a general 5D supersymmetric Yang-Mills (SYM) theory with 8 supersymmetries, by choosing the space-like vector vev $\langle C^\mu\rangle=g^2_{\rm YM}\d^\mu_5$. Here $C^\mu$ is an auxiliary field, and $g_{\rm YM}$ the coupling constant of the supersymmetric Yang-Mills theory.
In Section \ref{SYM}, we discuss some other cases with $\langle C^\mu\rangle$ being a light-like or a time-like vector. It would be interesting to investigate these SYM theories.


Our paper is organized as follows. In Section \ref{secmini}, we review the  ``minimal" $(1,0)$ tensor multiplet theory of our previous work \cite{Chen10}; In Section \ref{sec1scft}, we  construct the 6D $(1,0)$ theory by coupling a $(1,0)$ hypermultiplet theory to this $(1,0)$ tensor multiplet theory. In Section \ref{SecLP}, we derive the $(2,0)$ LP theory by enhancing the supersymmetry from $(1,0)$ to $(2,0)$. In Section \ref{SYM}, we construct the action of the $\CN=1$ SYM theory in 5D, by setting $\langle C^\mu\rangle=g^2_{\rm YM}\d^\mu_5$ in the $(1,0)$ theory; We also briefly discuss the applications of these theories to M5-branes. 
In Appendix \ref{Mini10}, we verify the closure of the superalgebra of the minimal $(1,0)$ tensor multiple theory. In Appendix \ref{secsveom}, we prove that the set of equations of motion of the 6D $(1,0)$ theory are closed under supersymmetry transformations. In Appendix \ref{secsupercurrents}, we construct the conserved supercurrents and discuss the possibilities for enhancing the Poincare supersymmetries to the full superconformal symmetries.

\section{Review of the Minimal $(1,0)$ Tensor Multiplet}\label{secmini}

In this section, we first review the 6D  nonabelian $(1,0)$ tensor multiplet theory\footnote{We also call it a ``minimal" $(1,0)$ tensor multiplet theory. After coupling to the $(1,0)$ hypermultiplet theory, it will be called a nonabelian $(1,0)$ theory.} constructed in Section 2 of \cite{Chen10}. We then recast it such that the $SU(2)$ R-symmetry is manifest.

\subsection{Review of  $(1,0)$ Tensor Multiplet}
Following the convention of \cite{Chen10}, we will first work with 32-component Majorana fermions. (More precisely, we will work with $SO(9,1)$ Majorana fermions.) The gamma matrices satisfy the anti-commutation relations
\e
\{\Gamma^\mu,\Gamma^\nu\}&=&2\eta^{\mu\nu},\quad (\mu=0, 1,\ldots,5.)\nonumber \\
\{\Gamma^s,\Gamma^t\}&=&2\d^{st},\quad (s, t=6, 7, 8, 9.)\nonumber\\
\{\Gamma^s,\Gamma^\mu\}&=&0,\label{gamma0}
\ee
where $\eta^{\mu\nu}={\rm diag}(-1,1,1,1,1,1)$. We begin by reviewing the free $(1, 0)$ theory of tensor multiplet. It contains a scalar field $\phi$, an antisymmetric form field $B_{\nu\rho}$, and a fermionic field
$\chi$.
The fermionic field $\chi$ is anti-chiral with respect to $\Gamma_{012345}$, but chiral with respect to $\Gamma_{6789}$:
\e\label{chiral5}
\Gamma_{012345}\chi&=&-\chi,\nonumber\\
\Gamma_{6789}\chi&=&\chi.
\ee
The above two equations imply that $\Gamma_{0123456789}\chi=-\chi$, i.e.
$\chi$ is a Weyl spinor. Recall that we assumed that $\chi$ is an $SO(9,1)$ Majorana spinor, so $\chi$ is an $SO(9,1)$ Majorana-Weyl spinor.

The supersymmetry transformations are
\e\label{susy1}
\d\phi&=&-i\bar\ep\chi,\nonumber\\
\d\chi&=&\Gamma^\mu\ep\partial_\mu\phi+\frac{1}{3!}\frac{1}{2!}
\Gamma_{\mu\nu\lambda}\ep H^{\mu\nu\lambda},\nonumber\\
\d H_{\mu\nu\rho}&=&
3i\bar\ep\Gamma_{[\mu\nu}\partial_{\rho]}\chi.
\ee
The self-dual field strength is defined as $H_{\mu\nu\rho}=3\partial_{[\mu}B_{\nu\rho]}$. The supersymmetry parameter\footnote{In Ref. \cite{Chen10}, the supersymmetry parameter is denoted as $\ep_+$, which is a 10D Majorana-Weyl spinor.} $\ep$
is chiral with respect to $\Gamma_{012345}$ as well as $\Gamma_{6789}$, i.e.,
\e
\Gamma_{012345}\ep&=&\ep,\nonumber\\
\Gamma_{6789}\ep&=&\ep.
\ee
The super-poincare algebra is closed by imposing the equations of motion (EOM)
\e
\Gamma^\mu\partial_\mu\chi=0,\quad\quad
\partial^\mu\partial_\mu\phi=0,\quad\quad\partial_{[\mu}H_{\nu\rho\s]}=0.
\ee
However, due to the self-duality nature of $H_{\mu\nu\rho}$, it is difficult to construct a Lagrangian. The reason is as follows: The kinetic term $H_{\mu\nu\rho}H^{\mu\nu\rho}$ is proportional to
\e
\vp^{\mu\nu\rho\s\lambda\tau}H_{\mu\nu\rho}H_{\s\lambda\tau},
\ee
which vanishes by the self-duality conditions. Here $\vp^{012345}=-\vp_{012345}=1$.

One can generalize the above free $(1,0)$ tensor multiplet to be the nonabelian one \cite{Chen10},
\e
(\phi_m, H_{\mu\nu\rho m},\chi_{m}).
\ee
(In Ref. \cite{Chen10}, the fermionic field  is denoted as $\psi_{m+}$.) Here $m$ is an adjoint index of the Lie algebra of gauge group, and $k_{mn}$ is an invariant form on the Lie algebra. If the Lie algebra is semi-simple, then $k_{mn}$ is nothing but the Killing-Cartan metric, whose inverse will be denoted as $k^{mn}$. We will use $k_{mn}$ to lower indices, and use its inverse $k^{mn}$ to raise indices; for instance, $\phi^m=k^{mn}\phi_n$.

The the components of the field strength $H_{\mu\nu\rho m}$ also obey the self-dual conditions
\e\label{selfdual0}
H_{\mu\nu\rho m}=\frac{1}{3!}\vp_{\mu\nu\rho\sigma\lambda\kappa}H^{\sigma\lambda\kappa}_m.
\ee
After introducing the nonabelian gauge symmetry, the law of supersymmetry reads \cite{Chen10}:
\e\label{susy2}
\d\phi_m&=&-i\bar\ep\chi_{m},\nonumber\\
\d\chi_{m}&=&\Gamma^\mu\ep D_\mu\phi_m+\frac{1}{3!}\frac{1}{2!}
\Gamma_{\mu\nu\lambda}\ep H^{\mu\nu\lambda}_m,
\nonumber\\
\d A_\mu^m&=&i\bar\ep\Gamma_{\mu\nu}\chi^{m}C^{\nu},
\nonumber\\
\d C^\nu&=&0,\nonumber\\
\d H_{\mu\nu\rho m}&=&
3i\bar\ep\Gamma_{[\mu\nu}D_{\rho]}\chi_{m}-i\bar\ep\Gamma_{\mu\nu\rho\sigma}
C^{\sigma}\chi_n\phi_pf^{np}{}_{m},
\ee
where $C^\mu$ is an abelian auxiliary field, and $f^{np}{}_{m}$ the structure constants of the Lie algebra of the gauge group. The covariant derivative is defined as follows
\e\label{covd}
D_\mu\phi_m=\partial_\mu\phi_m+ (A_\mu)_n\phi_pf^{np}{}_m.
\ee
The equations of the nonabelian $(1,0)$ theory are given by \cite{Chen10}
\e\label{eqs1}
0&=&D^2\phi_p-\frac{i}{2}(\bar\chi_{m}\Gamma_\nu\chi_{n})C^\nu f^{mn}{}_p,
\nonumber\\
0&=&F^m_{\mu\nu}-H^m_{\mu\nu\rho}C^\rho,\nonumber\\
0&=&\Gamma^\mu D_\mu\chi_{m}-\Gamma^\mu C_\mu\chi_{n}\phi_pf^{np}{}_m,
\nonumber\\
0&=&D_{[\mu}H_{\nu\rho\s]p}+\frac{i}{8}\vp_{\mu\nu\rho\lambda\s\tau}(\bar\chi_{m}\Gamma^\tau
\chi_{n})C^\lambda f^{mn}{}_p+\frac{1}{4}\vp_{\mu\nu\rho\lambda\s\tau}\phi_mC^\lambda D^\tau\phi_n f^{mn}{}_p,
\nonumber\\
0&=&C^\s D_\s\phi^{m}=C^\s D_\s\chi^{m}=C^\s D_\s H_{\mu\nu\rho }^m
=C^\s D_\s F_{\mu\nu}^m=\partial_\mu C^\nu.
\ee
The field strength $F^m_{\mu\nu}$ is defined as
\e
F^m_{\mu\nu}&=&\partial_\mu A^m_\nu-\partial_\nu A^m_\mu+[A_\mu,A_\nu]^m.
\ee
The supersymmetry transformations (\ref{susy2}) are closed, provided that the equations (\ref{eqs1}) are obeyed.

\subsection{ $(1,0)$ Tensor Multiplet with Manifest $SU(2)$ R-symmetry}\label{secsmall}
In this section, we recast the theory such that the $SU(2)$ R-symmetry is manifest.
Notice that the $SO(9,1)$ $\Gamma$-matrices can be constructed as follows
\e
(\Gamma^\mu)_{\rm10D}&=&(\Gamma^\mu)_{\rm 6D}\otimes(\gamma^5)_{\rm 4D}, \quad \mu=0, 1,\ldots,5.\nonumber\\
(\Gamma^s)_{\rm10D}&=&\textbf{1}_{8\times8}\otimes(\gamma^s)_{\rm 4D},\quad s=6, 7, 8, 9.\label{gamma1}
\ee
Here $(\gamma^s)_{\rm 4D}$ are the set of $SO(4)=SU(2)\times SU(2)$ hermitian matrices
\e\label{4dgamma}
\gamma^s=\left(
  \begin{array}{cc}
    0 & \s^s \\
   \s^{s\dag} & 0 \\
  \end{array}
\right),\label{so4}
\ee
where $\s^s=(\vec{\s},i\textbf{1}_{2\times2})$ and $\s^{s\dag}=(\vec{\s}, -i\textbf{1}_{2\times2})$, with $\vec{\s}$ the pauli matrices.
And
\e\label{gamma5}
(\gamma^5)_{\rm 4D}=(\gamma^6\gamma^7\gamma^8\gamma^9)_{\rm 4D}=\left(
  \begin{array}{cc}
    \d^A_B & 0 \\
    0 & -\d^\DA_\DB\\
  \end{array}
\right),
\ee
where $A, B=1, 2$ and $\DA, \DB=\dot1, \dot2$ are the undotted and dotted indices of $SU(2)\times SU(2)$, respectively.
The $8\times8$ gamma matrices $(\Gamma^\mu)_{\rm 6D}$ are defined as
\e\label{6dgamma}
(\Gamma^\mu)_{\rm 6D}=\left(
  \begin{array}{cc}
    0 & \tilde\Gamma^\mu \\
    \Gamma^\mu & 0 \\
  \end{array}
\right).
\ee
In the right hand side, $\Gamma^\mu$ can be chosen as the set of $4\times4$ matrices
\e\label{gamma6}
\Gamma^0&=&-\s^3\otimes\s^2,\quad \Gamma^1=\s^2\otimes\s^3,\quad\Gamma^2=i\s^2\otimes\textbf{1}_{2\times2},\nonumber\\
\Gamma^3&=&\s^2\otimes\s^1,\quad \Gamma^4=i\s^1\otimes\s^2,\quad \Gamma^5=\textbf{1}_{2\times2}\otimes\s^2,
\ee
and $\tilde\Gamma^\mu=\Gamma^\dag_\mu$, satisfying
\e
\Gamma^\mu\tilde\Gamma^\nu+\Gamma^\nu\tilde\Gamma^\mu=2\eta^{\mu\nu}\quad{\rm and}\quad \tilde\Gamma^\mu\Gamma^\nu+\tilde\Gamma^\nu\Gamma^\mu=2\eta^{\mu\nu}.\label{so6}
\ee
Using equations (\ref{so4})$-$(\ref{so6}), we see that (\ref{gamma1}) indeed satisfy
the commutation relations (\ref{gamma0}).

Equations (\ref{gamma1}) are
essentially the decomposition: $SO(9,1)\Rightarrow SO(5,1)\times SU(2)\times SU(2)$. 
Equations (\ref{gamma0}), (\ref{chiral5}), and (\ref{gamma1}) suggest that the $SO(9,1)$ Majorana-Weyl fermion $\chi_{\Sigma}$ can be converted into an $SU(2)$ symplectic-Majorana chiral spinor $\chi_{\a A}$:
\e\label{decom1}
&\chi_{\Sigma}\rightarrow& \chi_{\a A},
\ee
where  $\Sigma$ labels the Majorana-Weyl representation of $SO(9,1)$, and $\a$ labels the Weyl representation of $SO(5,1)$, more precisely,
\e
(\Gamma_{012345})_{\rm 6D}\chi_{A}=-\chi_{A},
\ee
and $A=1,2$ is a fundamental index of the $SU(2)$ R-symmetry group.

In the basis (\ref{gamma1}), the reality condition (Majorana condition) reads
\e\label{reality1}
(\chi_{m})^*=B_{\rm10D}\chi_{m}\quad,
\ee
where
\e\label{B}
B_{\rm10D}
&=&B_{\rm6D}\otimes B_{\rm4D},\nonumber\\
B_{\rm6D}&=&\left(
  \begin{array}{cc}
    \overline{B}_{\rm4D} & 0 \\
    0 & B_{\rm4D} \\
  \end{array}
\right)
=\left(
  \begin{array}{cc}
    \s^3\otimes i\s^2 & 0 \\
    0 & -\s^3\otimes i\s^2 \\
  \end{array}
\right),
\nonumber\\
B_{\rm4D}&=&\left(
  \begin{array}{cc}
    \ep^{AB} & 0 \\
    0 & \ep^{\DA\DB} \\
  \end{array}
\right)
=\left(
  \begin{array}{cc}
    -i\s^2 & 0 \\
    0 & i\s^2 \\
  \end{array}
\right).
\ee
We denote the inverses of the anti-symmetric forms $\ep^{\DA\DB}$ and $\ep^{AB}$ as $\ep_{\DB\DC}$ and $\ep_{BC}$, respectively, satisfying $\ep^{\DA\DB}\ep_{\DB\DC}=\d^\DA_\DC$ and $\ep^{AB}\ep_{BC}=\d^A_C$.  Now the reality condition (\ref{reality1}) is equivalent to  $SU(2)$ Majorana condtion
\e\label{reality3}
(\chi_{Am})^*=\ep^{AB}B_{\rm6D}\chi_{Bm}.
\ee
Similarly, the 10D Majorana-Weyl spinor $\ep_{\Sigma}$ can be converted into the $SU(2)$ simplectic Majorana spinor $\ep_{\a A}$, i.e. $\ep_{\Sigma}\rightarrow \ep_{\a A}$. Here $\ep_{A}$ obeys the reality and chirality conditions:
\e\label{reality2}
&&(\Gamma_{012345})_{\rm 6D}\ep_{A}=\ep_{ A},\nonumber\\
 &&(\ep_{A})^*=\ep^{AB}B_{\rm6D}\ep_{B}.\label{reality2}
\ee

Using (\ref{gamma1}), (\ref{decom1}), (\ref{reality3}), and (\ref{reality3}), the law of supersymmetry transformation (\ref{susy2}) can be recast into the form\footnote{In Appendix \ref{secsveom}, the super-variation ``$\d$" in (\ref{susy3}) will be replaced by  ``$\bar\d$", while the super-variation in (\ref{susy5}) will be still denoted as ``$\d$".}
\e\label{susy3}
\d\phi_m&=&-i\bar\ep^A\chi_{Am},\nonumber\\
\d\chi_{Am}&=&\Gamma^\mu\ep_AD_\mu\phi_m+\frac{1}{3!}\frac{1}{2!}
\Gamma_{\mu\nu\lambda}\ep_AH^{\mu\nu\lambda}_m,
\nonumber\\
\d A_\mu^m&=&i\bar\ep^A\Gamma_{\mu\nu}\chi^m_{A}C^{\nu},
\nonumber\\
\d C^\nu&=&0,\nonumber\\
\d H_{\mu\nu\rho m}&=&
3i\bar\ep^{A}\Gamma_{[\mu\nu}D_{\rho]}\chi_{Am}-i\bar\ep^{A}\Gamma_{\mu\nu\rho\sigma}
C^{\sigma}\chi_{An}\phi_pf^{np}{}_{m},
\ee
and the equations (\ref{eqs1}) can be recast into
\e\label{eqs2}
0&=&D^2\phi_p-\frac{i}{2}(\bar\chi^A_{m}\Gamma_\nu\chi_{An})C^\nu f^{mn}{}_p,
\nonumber\\
0&=&F^m_{\mu\nu}-H^m_{\mu\nu\rho}C^\rho,\nonumber\\
0&=&\Gamma^\mu D_\mu\chi_{Am}-\Gamma^\mu C_\mu\chi_{An}\phi_pf^{np}{}_m,
\nonumber\\
0&=&D_{[\mu}H_{\nu\rho\s]p}+\frac{i}{8}\vp_{\mu\nu\rho\lambda\s\tau}
(\bar\chi^A_{m}\Gamma^\tau
\chi_{An})C^\lambda f^{mn}{}_p+\frac{1}{4}\vp_{\mu\nu\rho\lambda\s\tau}\phi_mD^\tau\phi_nC^\lambda f^{mn}{}_p,
\nonumber\\
0&=&C^\s D_\s\phi_{m}=C^\s D_\s\chi_{Am}=C^\s D_\s H_{\mu\nu\rho m}=\partial_\mu C^\nu,
\ee
where the gamma matrices in (\ref{susy3}) and (\ref{eqs2}) are defined by (\ref{6dgamma}), 
and we have dropped the subscript ``6D", i.e.
\e
(\Gamma^\mu)_{6D}\rightarrow \Gamma^\mu.
\ee
It can be seen that in (\ref{susy3}) and (\ref{eqs2}), the $SU(2)$ R-symmetry is manifest. In Appendix \ref{Mini10}, we rederive equations (\ref{eqs2}) by requiring the closure of the super Poincare algebra.

It is well know that the gauge field of the $\CN=6$ ABJM theory \cite{ABJM} is non-dynamical. Here the gauge field $A^m_\mu$ is also  non-dynamical. If it were a dynamical field,
its super-partner (gaugino)  would be also an independent dynamical field.
However, the third equation
of (\ref{susy3}) indicates that the gaugino can be expressed in term of the
 fermionic field $\chi^m$ of the tensor multiplet and the auxiliary field $C^\mu$. So the gaugino is just an auxiliary field. In other words, the gaugino is non-dynamical.

\section{Nonabelian $(1,0)$ Theory}\label{sec1scft}
In this section, we will construct the nonabelian $\CN=(1,0)$ theory by coupling the  $\CN=(1,0)$ tensor multiplet theory to an $\CN=(1,0)$ hypermultiplet theory.
\subsection{Closure of the $\CN=(1,0)$ Superalgebra}
We begin by presenting a quick review of the free theory of hypermultiplet. The supersymmetry transformations are given by
\e
\d\phi^A&=& i\bar\ep^A\psi,\nonumber\\
\d\psi&=& -2\ep_A\Gamma^\mu \partial_\mu \phi^A.
\ee
Here $\ep_A$ satisfies the reality and chirality conditions (\ref{reality2}), and $A=1,2$ is a fundamental index of the R-symmetry group $SU(2)$. The fermionic field $\psi$ is a 6D Weyl spinor, and it is anti-chiral with respect the 6D chirality matrix, i.e.
\e\label{psichiral}
\Gamma_{012345}\psi=-\psi.
\ee
The super-Poincare algebra is closed provided the equations of motion
\e
\Gamma^\mu\partial_\mu\psi=0\quad{\rm and}\quad \partial^\mu\partial_\mu\phi^A=0
\ee
are satisfied.

To couple the hypermultiplet and the  tensor multiplet, it is natural to assume that they share the same gauge symmetry. Recall that the tensor multiplet constructed in the last section is in the adjoint representation of the Lie algebra of gauge group. However, it is not necessary to assume that the hypermultiplet is also in the adjoint representation. \emph{Instead, we assume that the hypermultiplet can be in the arbitrary representation of the gauge group; in particular, it can be in the fundamental representation of the gauge group}\footnote{We emphasize this point because the matter fields of the $\CN=6$ ABJM theory are also in the bi-fundamental representation of the gauge group $U(N)\times U(N)$. In fact, to achieve enhanced supersymmetries ($\CN\geq4$), the Lie algebras of gauge groups of 3D Chern-Simons matter theories must be chosen as the bosonic parts of certain superalgebras, and the matter fields matter fields must be in the fundamental representations of these Lie algebras.
However, here the Lie algebra of the gauge group of the $(1,0)$ theory can be arbitrary, not necessarily restricted to the bosonic part of some superalgebra. It would be interesting to study the Lie algebras of gauge groups and the corresponding representations for both 3D and 6D theories.}. With this understanding, the component fields of the nonabelian hypermultiplet can be written as
\e
(\phi^A_I, \psi_I),
\ee
where $I$ labels an arbitrary representation of the Lie algebra of gauge symmetry. The complex conjugation of $\phi^A_I$ will be denoted as $\bar\phi^I_A$, i.e. $\bar\phi^I_A=(\phi^A_I)^*$. The covariant derivative is defined as
\e
D_\mu\phi^A_I=\partial_{\mu}\phi^A_I-\tau^{mJ}{}_IA_{\mu m}\phi^A_J,
\ee
where $\tau^{mJ}{}_I$ are a set of representation matrices of the generators of the gauge group, and $A_{\mu m}=k_{mn}A^n_\mu$. To ensure the positivity of the theory, we assume that $\tau^{mJ}{}_I$ obeys the reality condition:
\e\label{tau1}
(\tau^{mJ}{}_I)^*=-\tau^{mI}{}_J.
\ee

We postulate the law of supersymmetry transformations as follows
\e\label{susy4}
\d\phi_m&=&-i\bar\ep^A\chi_{Am},\nonumber\\
\d\phi^A_I&=&i\bar\ep^A\psi_I,\nonumber\\
\d\chi_{Am}&=&\Gamma^\mu\ep_AD_\mu\phi_m+\frac{1}{3!}\frac{1}{2!}\Gamma_{\mu\nu\lambda}\ep_AH^{\mu\nu\lambda}_m+a_1\Gamma_\lambda\ep_BC^\lambda(\bar\phi^J_A\phi^B_I+\bar\phi^{BJ}\phi_{AI})\tau_m{}^I{}_J,
\nonumber\\
\d\psi_I&=& -2\ep_A\Gamma^\mu D_\mu \phi^A_I+b_1\Gamma_\lambda\ep_AC^\lambda\tau^{mJ}{}_I\phi_m\phi^A_J,\nonumber\\
\d A_\mu^m&=&i\bar\ep^A\Gamma_{\mu\nu}\chi^m_{A}C^{\nu},
\nonumber\\
\d C^\nu&=&0,\nonumber\\
\d H_{\mu\nu\rho m}&=&
3i\bar\ep^{A}\Gamma_{[\mu\nu}D_{\rho]}\chi_{Am}-i\bar\ep^{A}\Gamma_{\mu\nu\rho\sigma}
\chi_{An}C^{\sigma}\phi_pf^{np}{}_{m}\nonumber\\
&&+id_1\bar\ep^A\Gamma_{\mu\nu\rho\s}\psi_IC^\s
\bar\phi^J_A\tau_m{}^I{}_J+id_2\bar\psi^I\Gamma_{\mu\nu\rho\s}\ep_AC^\s\phi^A_J\tau_m{}^J{}_I,
\ee
where $\bar\phi^{BJ}=\vp^{BA}\bar\phi^J_A$ and $\phi_{AI}=\ep_{AB}\phi^{B}_I$, and $a_1$, $b_1$, $d_1$, and $d_2$ are real constants, to be determined later.

We now check the closure of the super-Poincare algebra. The supersymmetry transformation of the scalar field $\phi_m$ is
\e\label{scalarten}
[\d_1,\d_2]\phi_m=v^\mu D_\mu\phi_m,
\ee
where
\e\label{v}
v^\mu\equiv-2i\bar\ep^A_2\Gamma^\mu\ep_{1A}.
\ee

The transformation on the scalar field $\phi^A_I$ is
\e\label{scalarhy}
[\d_1,\d_2]\phi^A_I=v^\mu D_\mu\phi^A_I+\frac{b_1}{2}\Lambda_m\tau^{mJ}{}_I\phi^A_J,
\ee
where
\e
\Lambda_m\equiv-v^\mu C_\mu\phi_m,
\ee
with $v^\mu$ defined by (\ref{v}). Later we will see that the second term of the right-hand side of (\ref{scalarhy}) is a gauge transformation.

Let us now look at the gauge field:
\e\label{gauge}
[\d_1,\d_2]A^m_\mu&=& v^\nu F^m_{\nu\mu}-D_\mu\Lambda^m\nonumber\\
&&+v^\nu(F^m_{\mu\nu}-H^m_{\mu\nu\rho}C^\rho)
\nonumber\\
&&-v_\mu(C^\nu D_\nu\phi^m).
\ee
We see that the second term of the first line is a gauge transformation by the parameter $\Lambda^m$. Requiring the second term of (\ref{scalarhy}) to be a gauge transformation determines the constant $b_1$:
\e
b_1=-2.
\ee
Also, since $[\Lambda,\phi]_m=0$, Eq. (\ref{scalarhy}) can be written in the desired form:
\e\label{scalarten2}
[\d_1,\d_2]\phi_m=v^\mu D_\mu\phi_m+[\Lambda,\phi]_m
\ee
To close the super-poincare algebra on the gauge field, we must require the last two lines of (\ref{gauge}) to vanish separately. This determines the equations of motion for the gauge fields
\e\label{eqsA}
0&=&F^m_{\mu\nu}-H^m_{\mu\nu\rho}C^\rho
\ee
and the constraint equation on the scalar fields $\phi^m$:
\e
0&=&C^\nu D_\nu\phi^m.
\ee
Taking a super-variation on the above equation gives
\e
0&=&C^\nu D_\nu\chi^m_A.
\ee

The supersymmetry transformation of the fermionic field $\psi_I$ is given by
\e\label{psi}
[\d_1,\d_2]\psi_I&=&v^\mu D_\mu\psi_I-\Lambda^{J}{}_I\psi_J\nonumber\\
&&-\frac{1}{2}v_\nu\Gamma^\nu\bigg(\Gamma^\mu D_\mu\psi_I+\Gamma^\mu C_\mu\tau^{mJ}{}_I\phi_m\psi_J-2\Gamma^\mu C_\mu\psi_{Am}\tau^{mJ}{}_I\phi^A_J\bigg),
\ee
where $\Lambda^{J}{}_I=\Lambda_m\tau^{mJ}{}_I$.
The first line of (\ref{psi}) is the translation and the gauge transformation. So the second line must be the equations of motion
\e\label{eompsi}
0=\Gamma^\mu D_\mu\psi_I+\Gamma^\mu C_\mu\tau^{mJ}{}_I\phi_m\psi_J-2\Gamma^\mu C_\mu\tau^{mJ}{}_I\phi^A_J.
\ee
In deriving (\ref{psi}), we have used the Fierz identity
\e\label{Fierz1}
\ep_{1A}\bar\ep^B_2=-\frac{1}{4}\bigg(\bar\ep^B_2\Gamma_\mu\ep_{1A}\bigg)
\Gamma^\mu\frac{1-\Gamma}{2}+\frac{1}{48}\bigg(\bar\ep^B_2\Gamma_{\mu\nu\rho-}\ep_{1A}\bigg)
\Gamma^{\mu\nu\rho}_+.
\ee
Here $\Gamma=\Gamma_{012345}$ is the chirality matrix of $SO(5,1)$; and
\e
\Gamma_{\mu\nu\rho\pm}\equiv \frac{1}{2}(\Gamma_{\mu\nu\rho}\pm\frac{1}{3!}
\ep_{\mu\nu\rho\s\lambda\tau}\Gamma^{\s\lambda\tau}),
\ee
obeying the duality conditions
\e
\Gamma_{\mu\nu\rho\pm}=\pm\frac{1}{3!}\ep_{\mu\nu\rho\s\lambda\tau}\Gamma^{\s\lambda\tau}_\pm.\label{duality}
\ee
The above two equations are special cases of the identity
\e\label{g1}
\Gamma_{\mu_1\ldots\mu_p}=\frac{(-1)^{\frac{1}{2}(p-1)p}}{(6-p)!}\vp_{\mu_1\ldots\mu_{p}\mu_{p+1}\ldots\mu_6}
\Gamma^{\mu_{p+1}\ldots\mu_6}\Gamma_{012345}.
\ee

The transformation on the fermionic fields $\chi_{Am}$ is given by
\e\label{chi1}
[\d_1,\d_2]\chi_{Am}&=&v^\mu D_\mu\chi_{Am}+[\Lambda,\chi_A]_m\nonumber\\
&&-\frac{1}{4}v_\nu\Gamma^\nu\bigg(\Gamma^\mu D_\mu\chi_{Am}+\Gamma^\mu C_\mu[\phi, \chi_A]_m\bigg)\nonumber\\
&&+3(-a_1+d_2)v^{\mu\nu\rho}_{(AB)}\Gamma_{\nu\rho}B^{-1}\psi^IC_\mu\phi^B_J\tau_m{}^J{}_I
\nonumber\\
&&+3(a_1+d_1)v^{\mu\nu\rho}_{(AB)}\Gamma_{\nu\rho}\psi_IC_\mu\bar\phi^{BJ}\tau_m{}^I{}_J
\nonumber\\
&&+\frac{1}{8}(3a_1-5d_2)v^\mu C_\mu B^{-1}\psi^I\phi_{AJ}\tau_m{}^J{}_I
\nonumber\\
&&+\frac{1}{8}(3a_1-d_2)v^\nu C^\mu\Gamma_{\mu\nu} B^{-1}\psi^I\phi_{AJ}\tau_m{}^J{}_I
\nonumber\\
&&-\frac{1}{8}(3a_1+5d_1)v^\mu C_\mu\psi_I\bar\phi^J_{A}\tau_m{}^I{}_J
\nonumber\\
&&-\frac{1}{8}(3a_1+d_1)v^\nu C^\mu\Gamma_{\mu\nu} \psi_I\bar\phi^J_{A}\tau_m{}^I{}_J,
\ee
where
\e
v^{\mu\nu\rho}_{(AB)}\equiv -\frac{i}{24}\bigg(\bar\ep_{2A}\Gamma^{\mu\nu\rho}_-\ep_{1B}
+\bar\ep_{2B}\Gamma^{\mu\nu\rho}_-\ep_{1A}\bigg),
\ee
with $\bar\ep_{2A}=\ep_{AB}\bar\ep_2^{B}$; and \e\psi^J=(\psi_J)^*;\ee and $B^{-1}$ is the inverse of $B$, defined by the second equation of (\ref{B}), with the subscript ``6D" omitted. The third and fourth lines of (\ref{chi1}) must vanish separately, since they contain the  set of unwanted parameters $v^{\mu\nu\rho}_{(AB)}$. We are thus led to
\e\label{ad}
d_2=a_1 \quad{\rm and}\quad d_1=-a_1.
\ee
Substituting (\ref{ad}) into (\ref{chi1}), a short calculation gives
\e\label{chi2}
&&[\d_1,\d_2]\chi_{Am}\nonumber\\
&=&v^\mu D_\mu\chi_{Am}+[\Lambda,\chi_A]_m\\
&&-\frac{1}{4}v_\nu\Gamma^\nu\bigg(\Gamma^\mu D_\mu\chi_{Am}+\Gamma^\mu C_\mu[\phi, \chi_A]_m+a_1\Gamma^\mu C_\mu B^{-1}\psi^{J}\phi_{AI}\tau_{m}{}^I{}_J
-a_1\Gamma^\mu C_\mu\psi_{J}\bar\phi^{I}_A\tau_{m}{}^J{}_I\bigg).\nonumber
\ee
The first line of (\ref{chi2}) is a covariant translation and a gauge transformation. In order to close the super-Poincare algebra, one must require the last line of (\ref{chi2}) to vanish,
\e\label{eomchi}
0=\Gamma^\mu D_\mu\chi_{Am}+\Gamma^\mu C_\mu[\phi, \chi_A]_m+a_1\Gamma^\mu C_\mu B^{-1}\psi^{J}\phi_{AI}\tau_{m}{}^I{}_J
-a_1\Gamma^\mu C_\mu\psi_{J}\bar\phi^{I}_A\tau_{m}{}^J{}_I,
\ee
which are the equations of motion of $\chi_{Am}$.

After some algebraic steps, we obtain the super-variation of the self-dual field strengths:
\e\label{tensor1}
&&[\d_1,\d_2]H_{\mu\nu\rho m}\nonumber\\
&=&v^\s D_\s H_{\mu\nu\rho m}+[\Lambda,H_{\mu\nu\rho}]_m\nonumber\\
&&+3(v_{[\mu}F_{\nu\rho]n}-v_{[\mu}H_{\nu\rho]\s n}C^\s)\phi_pf^{np}{}_m\\
&&+4v^\s\bigg[D_{[\mu}H_{\nu\rho\s]m}-\frac{i}{8}\vp_{\mu\nu\rho\s\lambda\tau}C^\lambda \bigg((\bar\chi^A_n\Gamma^\tau
\chi_{Ap}) f^{np}{}_m+a_1(\bar\psi^J\Gamma^\tau\psi_I)\tau_m{}^I{}_J\bigg)\nonumber\\
&&\quad\quad\quad-\frac{1}{4}\vp_{\mu\nu\rho\s\lambda\tau}C^\lambda\bigg(\phi_nD^\tau\phi_p f^{np}{}_m+a_1\bar\phi^J_BD^\tau\phi^B_I\tau_m{}^I{}_J
-a_1\phi^B_JD^\tau\bar\phi^I_B
\tau_m^J{}_I\bigg)\bigg]\nonumber\\
&&+2ia_1(\bar\ep^A_2\Gamma_{\mu\nu\rho}\ep_{1B}
-\bar\ep^A_1\Gamma_{\mu\nu\rho}\ep_{2B})(C^\lambda D_\lambda\phi^B_I)\bar\phi^J_A\tau_m{}^I{}_J\nonumber\\
&&
+2ia_1(\bar\ep^B_2\Gamma_{\mu\nu\rho}\ep_{1A}
-\bar\ep^B_1\Gamma_{\mu\nu\rho}\ep_{2A})(C^\lambda
D_\lambda\bar\phi^A_J)\phi^I_B\tau_m{}^J{}_I\nonumber
\ee
The second line vanishes by the equations of motion of the gauge fields (\ref{eqsA}); In order to close the superalgebra on $H_{\mu\nu\rho m}$, the last four lines must also vanish. This gives the constraint equations on the scalar fields
\e\label{cscalar}
0=C^\lambda D_\lambda\phi^B_I=C^\lambda D_\lambda\bar\phi_B^I,
\ee
and the equations of motion of $H_{\mu\nu\rho m}$:
\e\label{EOMH}
0&=&D_{[\mu}H_{\nu\rho\s]m}-\frac{i}{8}\vp_{\mu\nu\rho\s\lambda\tau}C^\lambda \bigg((\bar\chi^A_n\Gamma^\tau
\chi_{Ap}) f^{np}{}_m+a_1(\bar\psi^J\Gamma^\tau\psi_I)\tau_m{}^I{}_J\bigg)\nonumber\\
&&\quad\quad\quad-\frac{1}{4}\vp_{\mu\nu\rho\s\lambda\tau}C^\lambda\bigg(\phi_nD^\tau\phi_p f^{np}{}_m+a_1\bar\phi^J_BD^\tau\phi^B_I\tau_m{}^I{}_J
-a_1\phi^B_JD^\tau\bar\phi^I_B
\tau_m^J{}_I\bigg).
\ee
Taking super-variations on (\ref{cscalar}), one obtains
\e
0=C^\lambda D_\lambda\psi_I=C^\lambda D_\lambda\psi^I.
\ee
The Bianchi identity $D_{[\mu}F_{\nu\rho]m}=0$ and Eqs. (\ref{eqsA}) and (\ref{EOMH}) imply that
\e
C^\s D_\s H_{\mu\nu\rho m}=0.
\ee

The equations of motion of $\phi^A_I$ and $\phi_p$ can be derived by taking super-variations on Eqs. (\ref{eompsi}) and (\ref{eomchi}), respectively (for details, see Appendix \ref{secsveom}). They are given by
\e
0&=&D^2\phi^A_I-iC^\nu(\bar\chi^{A}_m\Gamma_\nu\psi_J)\tau^{mJ}{}_I
+C^2\phi^A_K\phi_m\phi_n\tau^{mK}{}_J\tau^{nJ}{}_I\nonumber\\
&&+a_1C^2(\bar\phi^K_B\phi^A_L+\bar\phi^{AK}\phi_{BL})\phi^B_J\tau_{m}{}^L{}_K\tau^{mJ}{}_I,
\\
0&=&D^2\phi_m-\frac{i}{2}C^\nu(\bar\chi^B_{n}\Gamma_\nu\chi_{Bp})f^{np}{}_m
+\frac{i}{2}a_1C^\nu(\bar\psi^J\Gamma_\nu\psi_I)\tau_m{}^I{}_J\nonumber\\
&&+2a_1C^2
\bar\phi^J_A\phi^A_I\phi^n(\tau_{(m}\tau_{n)})^I{}_J.
\ee

We see that $a_1$ cannot be fixed by the closure of superalgebra. However, if $a_1\neq0$, it can be  absorbed into the redefinitions of the hypermultiplet fields:
\e\label{red}
&&\sqrt{a_1}\phi^A_I\rightarrow \phi^A_I,\quad \sqrt{a_1}\bar\phi_A^I\rightarrow \bar\phi_A^I,\nonumber\\
&&\sqrt{a_1}\psi_I\rightarrow \psi_I,\quad \sqrt{a_1}\psi^I\rightarrow \psi^I.
\ee
One can of course keep this continuous free parameter $a_1$ in the $(1,0)$ theory. If $a_1=0$, the theory is reduced to the minimal $(1,0)$ tensor multiplet theory of Section \ref{secmini}. It would be interesting to investigate the physical meaning of this continuous free parameter $a_1$.


\subsection{Summary of the Nonabelian $(1,0)$ STheory}\label{secNe10}
In summary, the equations of the $(1,0)$ theory are given by
\e\label{eqs3}
0&=&D^2\phi^A_I-iC^\nu(\bar\chi^{A}_m\Gamma_\nu\psi_J)\tau^{mJ}{}_I
+C^2\phi^A_K\phi_m\phi_n\tau^{mK}{}_J\tau^{nJ}{}_I\nonumber\\
&&+C^2(\bar\phi^K_B\phi^A_L+\bar\phi^{AK}\phi_{BL})\phi^B_J\tau_{m}{}^L{}_K\tau^{mJ}{}_I,
\nonumber\\
0&=&D^2\phi_m-\frac{i}{2}C^\nu(\bar\chi^B_{n}\Gamma_\nu\chi_{Bp})f^{np}{}_m
+\frac{i}{2}C^\nu(\bar\psi^J\Gamma_\nu\psi_I)\tau_m{}^I{}_J\nonumber\\
&&+2C^2
\bar\phi^J_A\phi^A_I\phi^n(\tau_{(m}\tau_{n)})^I{}_J,
\nonumber\\
0&=&F^m_{\mu\nu}-H^m_{\mu\nu\rho}C^\rho,\nonumber\\
0&=&\Gamma^\mu D_\mu\psi_I+\Gamma^\mu C_\mu\tau^{mJ}{}_I\phi_m\psi_J-2\Gamma^\mu C_\mu\chi_{Am}\tau^{mJ}{}_I\phi^A_J,\\
0&=&\Gamma^\mu D_\mu\chi_{Am}+\Gamma^\mu C_\mu[\phi, \chi_A]_m+\Gamma^\mu C_\mu B^{-1}\psi^{J}\phi_{AI}\tau_{m}{}^I{}_J
-\Gamma^\mu C_\mu\psi_{J}\bar\phi^{I}_A\tau_{m}{}^J{}_I,\nonumber\\
0&=&D_{[\mu}H_{\nu\rho\s]m}-\frac{i}{8}\vp_{\mu\nu\rho\s\lambda\tau}C^\lambda \bigg((\bar\chi^A_n\Gamma^\tau
\chi_{Ap}) f^{np}{}_m+(\bar\psi^J\Gamma^\tau\psi_I)\tau_m{}^I{}_J\bigg)\nonumber\\
&&\quad\quad\quad-\frac{1}{4}\vp_{\mu\nu\rho\s\lambda\tau}C^\lambda\bigg(\phi_nD^\tau\phi_p f^{np}{}_m+\bar\phi^J_BD^\tau\phi^B_I\tau_m{}^I{}_J
-\phi^B_JD^\tau\bar\phi^I_B
\tau_m^J{}_I\bigg),
\nonumber\\
0&=&C^\s D_\s\phi^{m}=C^\s D_\s\phi_I^{A}
=C^\s D_\s\chi^{m}=C^\s D_\s\psi_{I}=C^\s D_\s H^m_{\mu\nu\rho }=C^\s D_\s F^m_{\mu\nu}=\partial_\mu C^\nu.\nonumber
\ee
Here $a_1$ has been absorbed into the redefinitions of the fields (see (\ref{red})). 
And the law of supersymmetry transformations are as follows
\e\label{susy5}
\d\phi_m&=&-i\bar\ep^A\chi_{Am},\nonumber\\
\d\phi^A_I&=&i\bar\ep^A\psi_I,\nonumber\\
\d\chi_{Am}&=&\Gamma^\mu\ep_AD_\mu\phi_m+\frac{1}{3!}\frac{1}{2!}\Gamma_{\mu\nu\lambda}\ep_AH^{\mu\nu\lambda}_m+\Gamma_\lambda\ep_BC^\lambda(\bar\phi^J_A\phi^B_I+\bar\phi^{BJ}\phi_{AI})\tau_m{}^I{}_J,
\nonumber\\
\d\psi_I&=& -2\Gamma^\mu\ep_A D_\mu \phi^A_I-2\Gamma_\lambda\ep_AC^\lambda\tau^{mJ}{}_I\phi_m\phi^A_J,\nonumber\\
\d A_\mu^m&=&i\bar\ep^A\Gamma_{\mu\nu}\chi^m_{A}C^{\nu},
\nonumber\\
\d C^\nu&=&0,\nonumber\\
\d H_{\mu\nu\rho m}&=&
3i\bar\ep^{A}\Gamma_{[\mu\nu}D_{\rho]}\chi_{Am}-i\bar\ep^{A}\Gamma_{\mu\nu\rho\sigma}
C^{\sigma}\chi_{An}\phi_pf^{np}{}_{m}\nonumber\\
&&-i\bar\ep^A\Gamma_{\mu\nu\rho\s}\psi_IC^\s
\bar\phi^J_A\tau_m{}^I{}_J+i\bar\psi^I\Gamma_{\mu\nu\rho\s}\ep_A\phi^A_J\tau_m{}^J{}_I.
\ee

We have verified that the set of equations (\ref{eqs3}) are closed under the supersymmetry transformations (\ref{susy5}): Taking a super-variation on any equation of (\ref{eqs3}) can transform it into some other equations of (\ref{eqs3}). For instance, if we take a super-variation on the first equation of (\ref{eqs3}) (the EOM of $\phi^A_I$), we will obtain the equations of motion of the gauge fields $A^m_\mu$, and spinor fields $\chi_{Am}$ and $\psi_{I}$. In other words, under supersymmetry transformations (\ref{susy5}),
\e
\d({\rm Any\ Equation})=0\label{danyeom}.
\ee
The details are presented in Appendix \ref{secsveom}.

It would be interesting to re-construct the theory using a superspace approach \cite{BP}.



\section{Enhancing to $(2,0)$ LP Theory}\label{SecLP}
In this section, we will promote the $(1,0)$ theory
to the $(2, 0)$ LP theory \cite{LP}. Recall the  $(1,0)$ tensor multiplet is in the adjoint representation of the gauge group, while the $(1,0)$  hypermultiplet can be in arbitrary representation. To promote the supersymmetry to $(2, 0)$, it is necessary that the  $(1,0)$ tensor multiplet and hypermultiplet are in the \emph{same representation} of the gauge group. We are therefore led to require that the hypermultiplet is also in the adjoint representation, i.e.
\e\label{adjoint}
\psi_I\rightarrow\psi_n,\quad\quad\quad
\phi^A_I\rightarrow\phi^A_n,
\ee
where $n$ is an adjoint index of the Lie algebra of the gauge group. Accordingly, the representation matrices should be the structure constants, i.e.
\e
\tau^{mI}{}_J&\rightarrow& (\tau^{m})^n{}_p\equiv f^{nm}{}_p,\label{adjoint2}
\ee
which also obey the reality condition
(\ref{tau1}).

Now we are ready to enhance the $SU(2)$ R-symmetry to $USp(4)\cong SO(5)$. To do so, let us define
\e
\psi_{\DA n}=-\frac{i}{\sqrt2}\left(
  \begin{array}{cc}
    -\psi_n \\
    B^{-1}\psi^*_n \\
  \end{array}
\right),
\ee
$B^{-1}$ is the inverse of $B$, and $B$ is defined by the second equation of (\ref{B}), with ``6D" omitted. It is not difficult to check that $\psi_{\DA n}$ obeys the reality conditions
\e\label{psi2}
\psi^*_{\DA n}=\ep^{\DA\DB}B\psi_{\DA n}.
\ee
Here $\ep^{\DA\DB}$ is defined by the last equation of (\ref{B}). It can be seen that $\psi_{\DA n}$ transforms in the dotted representation of $SU(2)\times SU(2)$. Now it is possible to combine $\psi_{\DA n}$ and $\chi_{\DA n}$
to form a \textbf{4} of $USp(4)\cong SO(5)$ \footnote{To avoid introducing too many indices, we still use the capital letters $A$, $B$, $\ldots$, to label the $USp(4)$
indices. We hope this will not cause any confusion.}:
\e
\psi_{An}=\left(
  \begin{array}{cc}
    \chi_{An} \\
    -\psi_{\DA n} \\
  \end{array}
\right),\label{usp4psi}
\ee
where in left hand side, $A=1,\ldots,4$ is a fundamental index of $USp(4)$; and in the right hand side, $A=1,2$ and $\DA=\dot 1,\dot 2$ are un-dotted and dotted index of $SU(2)\times SU(2)$, respectively; The reality conditions become
\e
(\psi_{An})^*&=&\omega^{AB}B\psi_{Bn},\\
\omega^{AB}&=&\left(
  \begin{array}{cc}
    \ep^{AB} & 0 \\
    0 & \ep^{\DA\DB} \\
  \end{array}
\right),\nonumber
\ee
where $\omega^{AB}$ is the invariant antisymmetric tensor of $USp(4)$.

The set of scalar fields of the hypermultiplet can be re-arranged such that they transform a \textbf{4} of $SO(4)$:
\e
\phi^s_n=-\frac{i}{\sqrt2}\bigg(\s^{s\dag}{}_{\dot 1}{}^A\bar\phi_{An}-\s^{s\dag}{}_{\dot 2}{}^A\ep_{AB}\phi^B_n\bigg)
\ee
where $\s^{s\dag}{}_\DA{}^A=(\vec\s, -i\textbf{1}_{2\times2})$, with $\vec\s$ the pauli matrices. And $\phi^s_n$ and $\phi_n$ can be combined to form a \textbf{5} of $SO(5)$:
\e
\phi^a_n=(\phi^s_n, \phi_n),
\ee
Here $a=1,\ldots,5$ is a fundamental index of $SO(5)$. Similarly, we use the matrices (\ref{4dgamma}) and (\ref{gamma5}) to define the set of $SO(5)$ gamma matrices
\e\label{so5gamma}
\gamma^a_A{}^B=(\gamma^s, \gamma^5),
\ee
where we have dropped the subscript ``4D".

Using equations (\ref{adjoint})$-$(\ref{so5gamma}), the equations of motion (\ref{eqs3}) can be recast into
\e\label{eqs4}
0&=&D^2\phi^a_p-\frac{i}{2}(\bar\psi^A_{m}\Gamma_\nu\gamma^a_A{}^B\psi_{Bn})C^\nu f^{mn}{}_p
-C^2\phi^b_m\phi^a_n\phi^b_qf^{mn}{}_of^{oq}{}_p,
\nonumber\\
0&=&F^m_{\mu\nu}-H^m_{\mu\nu\rho}C^\rho,\nonumber\\
0&=&\Gamma^\mu D_\mu\psi_{Am}-\Gamma^\mu\gamma^a_A{}^BC_\mu\psi_{Bn}\phi^a_pf^{np}{}_{m},\\
0&=&D_{[\mu}H_{\nu\rho\s]p}+\frac{i}{8}\vp_{\mu\nu\rho\lambda\s\tau}(\bar\psi^A_{m}
\Gamma^\tau
\psi_{An})C^\lambda f^{mn}{}_p+\frac{1}{4}\vp_{\mu\nu\rho\lambda\s\tau}\phi^a_mD^\tau \phi^a_nC^\lambda f^{mn}{}_p,
\nonumber\\
0&=&C^\s D_\s \phi^a_{m}
=C^\s D_\s\psi_{Am}=C^\s D_\s H_{\mu\nu\rho m}=\partial_\mu C^\nu,\nonumber
\ee
where $A=1, \ldots,4$ is a fundamental index of the R-symmetry group $USp(4)$. We now see that the  $USp(4)\cong SO(5)$ R-symmetry is manifest. These equations are essentially the same equations of motion of the $\CN=(2,0)$ LP theory, constructed in terms of Nambu 3-algebra \cite{LP}. If we introduce the notation
\e
\psi^{\pm}_{Am}&=&\frac{1}{2}(1\pm\gamma^5)_A{}^B\psi_{Bm},\nonumber\\
\psi_{Am}&=&\psi^{+}_{Am}+\psi^{-}_{Am},
\ee
we see that the equations of motion (\ref{eqs4}) are invariant if we switch $\psi^{+}_{Am}$ and $\psi^{-}_{Am}$:
\e
\psi^{+}_{Am}\leftrightarrow\psi^{-}_{Am}
\ee
Later we will see that the above \emph{discrete} symmetry allows us to enhance the $\CN=(1,0)$ supersymmetry to $\CN=(2,0)$. 

For convenience, we define two sets of parameters of supersymmetry transformations as follows:
\e
\ep_{A\pm}&=&\frac{1}{2}(1\pm\gamma^5)_A{}^B\ep_{B},\quad (A,\ B=1,\ldots,4.)
\ee
where
\e\label{usp4para}
\ep_{A}&=&\left(
  \begin{array}{cc}
    \ep_{A} \\
    \ep_{\DA } \\
  \end{array}
\right).
\ee
In the right hand side, $A=1,\ldots,4$ is a fundamental index of $USp(4)$, and the right hand side, $A=1, 2$ and $\DA=\dot1, \dot2$ are undotted and dotted index of $SU(2)\times SU(2)$.

Using equations (\ref{adjoint})$-$(\ref{so5gamma}), the supersymmetry transformations (\ref{susy5}) can be rewritten as
\e\label{susy6}
\d \phi^a_m&=&-i\bar\ep^A_+\gamma^a_A{}^B\psi_{Bm},\nonumber\\
\d\psi_{Am}&=&\Gamma^\mu\gamma^a_A{}^B\ep_{B+} D_\mu \phi^a_m+\frac{1}{3!}\frac{1}{2!}
\Gamma_{\mu\nu\lambda}\ep_{A+} H^{\mu\nu\lambda}_m+\frac{1}{2}\Gamma_\lambda\gamma^{ab}_A{}^B
\ep_{B+} C^{\lambda } \phi^a_n\phi^b_p f^{np}{}_{m},
\nonumber\\
\d A_\mu^m&=&i\bar\ep^A_+\Gamma_{\mu\nu}\psi^m_AC^{\nu},
\nonumber\\
\d C^\nu&=&0,\nonumber\\
\d H_{\mu\nu\rho m}&=&
3i\bar\ep^A_+\Gamma_{[\mu\nu}D_{\rho]}\psi_{Am}
-i\bar\ep^A_+\gamma^a_A{}^B\Gamma_{\mu\nu\rho\sigma}\psi_{Bn}C^{\sigma
}\phi^a_pf^{np}{}_{m}.
\ee
We see that in (\ref{susy6}), if we replace $\ep_{A+}$ by $\ep_{A-}$, while switch $\psi^{+}_{Am}$ and $\psi^{-}_{Am}$, that is,
\e
&&\ep_{A+}\rightarrow\ep_{A-}\nonumber\\
&&\psi^{+}_{Am}\leftrightarrow\psi^{-}_{Am}
\ee
we will obtain another \emph{independent }$\CN=(1, 0)$ supersymmetry transformations, whose R-symmetry is another $SU(2)$:
\e\label{susy7}
\d \phi^a_m&=&-i\bar\ep^A_-\gamma^a_A{}^B\psi_{Bm},\nonumber\\
\d\psi_{Am}&=&\Gamma^\mu\gamma^a_A{}^B\ep_{B-} D_\mu \phi^a_m+\frac{1}{3!}\frac{1}{2!}
\Gamma_{\mu\nu\lambda}\ep_{A-} H^{\mu\nu\lambda}_m+\frac{1}{2}\Gamma_\lambda\gamma^{ab}_A{}^B
\ep_{B-} C^{\lambda } \phi^a_n\phi^b_p f^{np}{}_{m},
\nonumber\\
\d A_\mu^m&=&i\bar\ep^A_-\Gamma_{\mu\nu}\psi^m_AC^{\nu},
\nonumber\\
\d C^\nu&=&0,\nonumber\\
\d H_{\mu\nu\rho m}&=&
3i\bar\ep^A_-\Gamma_{[\mu\nu}D_{\rho]}\psi_{Am}
-i\bar\ep^A_-\gamma^a_A{}^B\Gamma_{\mu\nu\rho\sigma}\psi_{Bn}C^{\sigma
}\phi^a_pf^{np}{}_{m}.
\ee
The equations of motion for closing the poincare supersymmetry algebra (\ref{susy7}) can be simply obtained by applying the discrete transformation $\psi^{+}_{Am}\leftrightarrow\psi^{-}_{Am}$ to (\ref{eqs4}). However, since $\psi^{+}_{Am}\leftrightarrow\psi^{-}_{Am}$ is just a discrete symmetry of (\ref{eqs4}). So the equations for closing (\ref{susy7}) are nothing but (\ref{eqs4}). In other words, the theory defined by (\ref{eqs4}) are invariant under the supersymmetry transformations (\ref{susy6}) \emph{and} (\ref{susy7}). Eqs. (\ref{susy6}) and (\ref{susy7}) can be unified to give the $\CN=(2, 0)$ supersymmetry transformations:
\e\label{susy8}
\d \phi^a_m&=&-i\bar\ep^A\gamma^a_A{}^B\psi_{Bm},\nonumber\\
\d\psi_{Am}&=&\Gamma^\mu\gamma^a_A{}^B\ep_B D_\mu \phi^a_m+\frac{1}{3!}\frac{1}{2!}
\Gamma_{\mu\nu\lambda}\ep_{A} H^{\mu\nu\lambda}_m+\frac{1}{2}\Gamma_\lambda\gamma^{ab}_A{}^B
\ep_{B} C^{\lambda } \phi^a_n\phi^b_p f^{np}{}_{m},
\nonumber\\
\d A_\mu^m&=&i\bar\ep^A\Gamma_{\mu\nu}\psi^m_AC^{\nu},
\nonumber\\
\d C^\nu&=&0,\nonumber\\
\d H_{\mu\nu\rho m}&=&
3i\bar\ep^A\Gamma_{[\mu\nu}D_{\rho]}\psi_{Am}
-i\bar\ep^A\gamma^a_A{}^B\Gamma_{\mu\nu\rho\sigma}\psi_{Bn}C^{\sigma
}\phi^a_pf^{np}{}_{m},
\ee
where $\ep_A$ is defined by (\ref{usp4para}). The above law of supersymmetry transformations is essentially the same as that of the $\CN=(2, 0)$ LP theory \cite{LP}. The above $(2, 0)$ supersymmetry transformations (\ref{susy8}) can be also obtained by re-casting the $(2, 0)$ supersymmetry transformations of \cite{Chen10}, using the gamma matrix decompositions in Section \ref{secsmall}. In enhancing the supersymmetry from $(1,0)$ to $(2,0)$, the Lie algebra of the gauge group of the theory can still be arbitrary, unlike the 3D $\CN\geq4$ superconformal Chern-Simons matter whose Lie algebras must be restricted to the bosonic parts of certain superalgebras.

In summary, Eqs. (\ref{eqs4}) and (\ref{susy8}), with manifest $USp(4)\cong SO(5)$ R-symmetry, are the ordinary Lie 2-algebra version of the $\CN=(2, 0)$ theory \cite{LP}.

\section{Relating to 5D SYM}\label{SYM}
In this section, we will demonstrate that upon dimension reduction, the 6D $\CN=(1, 0)$ theory in Section \ref{sec1scft} can be reduced to a general 5D $\CN=1$ SYM theory. Following the idea of Ref. \cite{LP}, we specify the space-like vector vev of $C^\mu$ as follows
\e\label{vev}
\langle C^\mu\rangle=g(0,\ldots,0,1)=g\d^\mu_5,
\ee
where the constant $g$ has dimension $-1$. Later we will see that it should be identified with $g^2_{\rm YM}$ \cite{LP}, i.e. $g=g_{\rm YM}$, where $g_{\rm YM}$ is the coupling constant of the 5D SYM theory.
Using (\ref{vev}), the equations of motion of gauge fields (the third equation of (\ref{eqs3})) are decomposed into
\e
&&F_{\a\b m}=gH_{\a\b5m},\nonumber\\
&&F_{5\b m}=gH_{5\b5m}=0,
\ee
where $\a, \b=0, 1,\ldots,4.$ The second equation says that
\e
F_{5\b}=\partial_5A_\b-\partial_\b A_5+[A_5,A_\b]=0
\ee
So $A_5$ is a flat connection. We may set $A_5=0$ at least locally, leading to
\e
\partial_5A_\b=0.
\ee
Namely, the gauge connection $A_\b$ is independent of the fifth coordinate $x^5$.
Also, substituting (\ref{vev}) into the last line of (\ref{eqs3}), we find that all other fields are also independent of the fifth coordinate $x^5$:
\e\label{no5}
0&=&\partial_5\phi_{m}=\partial_5 \phi^A_I
=\partial_5\chi_{Am}=\partial_5\psi_{I}=\partial_5 H_{\mu\nu\rho m}=\partial_5 g.
\ee


For convenience, we define the $SO(4, 1)$ gamma matrices as follows
\e\label{so5gamma2}
\gamma^\a\equiv i\Gamma^5\tilde\Gamma^\a,\quad (\a=0, 1, \ldots, 4.)
\ee
where the $\Gamma$-matrices are the $4\times4$ matrices defined by (\ref{gamma6}). Using (\ref{gamma6}), one can check that the set of gamma matrices $\gamma^\a$ obeys the Clifford algebra
\e
\{\gamma^\a, \gamma^\b\}=2\eta^{\a\b}.
\ee
Applying (\ref{vev}) to the rest equations of (\ref{eqs3}), and taking account of (\ref{no5}), it is natural to identify the 4-component Weyl spinor fields $(i\Gamma^5\chi_{Am})_{\rm6D}$ with the spinor fields $(\chi_{Am})_{\rm 5D}$.
(We have used ``6D" and ``5D" to indicate the dimensions of the corresponding spacetimes.) Specifically,
\e
&&i(\Gamma^5\chi_{Am})_{\rm 6D}=\left(
                         \begin{array}{cc}
                           0 & (\tilde\Gamma^5)_{4\times4} \\
                           (\Gamma^5)_{4\times4} & 0 \\
                         \end{array}
                       \right)\left(
                         \begin{array}{cc}
                           \chi_{Am} \\
                          0 \\
                         \end{array}
                       \right)
= (\chi_{Am})_{\rm 5D}.\nonumber
\ee
We have used
\e
\Gamma_{012345}(\chi_{Am})_{\rm6D}=-(\chi_{Am})_{\rm6D} \quad {\rm and} \quad \Gamma_{012345}=\left(
                         \begin{array}{cc}
                          -\textbf{1}_{4\times4} & 0 \\
                           0 & \textbf{1}_{4\times4} \\
                         \end{array}
                       \right),
\ee
where $(\Gamma^5)_{4\times4}$ is the gamma matrix defined in (\ref{gamma6}). Similarly, applying (\ref{vev}) and (\ref{no5}) to (\ref{eqs3}), it is possible to identify the 4-component Weyl spinor fields $i(\Gamma^5\psi_{I})_{\rm 6D}$ with the spinor fields $(\psi_{I})_{\rm 5D}$. In summary,
\e
&&i(\Gamma^5\chi_{Am})_{\rm 6D}=(\chi_{Am})_{\rm5D},\nonumber\\
&&i(\Gamma^5\psi_{I})_{\rm 6D}=(\psi_{I})_{\rm 5D}.
\ee
The above equations are also in accordance with (\ref{so5gamma2}). Without causing confusion, we will drop the subscript ``5D" of the spinor fields as we formulate the 5D SYM theory in the following paragraphs.

The 5D spinor fields $\chi_{Am}$ obey the reality conditions
\e\label{reality4}
(\chi_{Am})^*=\ep^{AB}\underline{B}\chi_{Bm},
\ee
where the $4\times 4$ matrix $\underline{B}$ is defined as
\e
\underline{B}=i\s^3\otimes\s^2.
\ee

Using (\ref{vev})$-$(\ref{reality4}), we are able to reduce the 6D equations (\ref{eqs3}) into the set of 5D equations of motion:
\e\label{5dEOM1}
0&=&\gamma^\a D_\a\psi_I-ig\psi_J\tau^{mJ}{}_I\phi_m+2ig\psi_{Am}\tau^{mJ}{}_I\phi^A_J,\nonumber\\
0&=&\gamma^\a D_\a\chi_{Am}-ig[\phi,\chi_A]_m-ig\underline{B}^{-1}\psi^J\phi_{AI}\tau_m{}^I{}_J
+ig\psi_I\bar\phi^J_A\tau_m{}^I{}_J,\nonumber\\
0&=&D^\a D_\a\phi_m-\frac{1}{2}g\bar\chi^A_n\chi_{Ap}f^{np}{}_m+\frac{1}{2}g\bar\psi^J
\psi_I\tau_m{}^I{}_J+2g^2\phi^n\bar\phi^{J}_A\phi^A_I(\tau_{(m}\tau_{n)})^I{}_J,\nonumber\\
0&=&D^\a D_\a\phi^A_I -g\bar\chi^{A}_m\psi_J\tau^{mJ}{}_I+g^2\phi^A_K\phi_m\phi_n\tau^{nK}{}_J\tau^{mJ}{}_I
\nonumber\\&&+g^2(\bar\phi^K_B\phi^A_L+\ep^{AC}\ep_{BD}\bar\phi^{K}_C\phi^D_{L})\phi^B_J
\tau_{m}{}^L{}_K\tau^{mJ}{}_I,
\nonumber\\
0&=&gD_{[\a}H_{\b\g]5m}=D_{[\a}F_{\b\g]m},
\nonumber\\
0&=&D^\a F_{\a\b p}-\frac{i}{2}g^2\bar\chi^A_m\gamma_\b\chi_{An}f^{mn}{}_p-\frac{i}{2}g^2\bar\psi^J
\gamma_\b\psi_I\tau_p{}^I{}_J\nonumber\\
&&-g^2\phi_mD_\b\phi_nf^{mn}{}_p-g^2(\bar\phi^J_AD_\b\phi^A_I
-\phi^A_ID_\b\bar\phi^J_A)\tau_p{}^I{}_J.
\ee
The covariant derivative $D_\a$ is defined as $D_\a=\partial_\a+A_\a$; for instance, $D_\a\phi_p=\partial_\a\phi_p+(A_\a)_m\phi_nf^{mn}{}_p.$ To formulate an action, we set
\e
g&=&g_{\rm YM},
\ee
and re-scale the fields as follows
\e
g\phi^I_A\rightarrow\phi^I_A,\quad\quad
g\psi_I\rightarrow\phi_I,\quad\quad
g\chi^A_m\rightarrow\chi^A_m,\quad\quad
g\phi^m\rightarrow\phi^m,
\ee
while leave the gauge field $A^\a$ unchanged, i.e. $A^\a\rightarrow A^\a$. The action of the 5D SYM theory with 8 supersymmetries is given by
\e\label{action}
L_{\rm YM}&=&\frac{1}{g^2_{\rm YM}}\bigg(\frac{1}{4}F^m_{\a\b}F^{\a\b}_m-\frac{i}{2}\bar\chi^{Am}
\gamma^\a D_\a\chi_{Am}-\frac{i}{2}\bar\psi^{I}
\gamma^\a D_\a\psi_I+\frac{1}{2}D^\a\phi^mD_\a\phi_m+D^\a\bar\phi^I_AD_\a\phi^A_I\nonumber\\
&&+\bar\psi^J\chi_{Am}\tau^{mI}{}_J\phi^A_I+\bar\chi^A_m\psi_{J}
\tau^{mJ}{}_I\bar\phi^J_A-\frac{1}{2}\bar\chi^{Am}\chi_{Ap}\phi_nf^{np}{}_m
-\frac{1}{2}\bar\psi^J\psi_I\phi_m\tau^{mI}{}_J\nonumber\\
&&-\phi_n\phi_m\bar\phi^J_A\phi^A_I\tau^{mI}{}_K\tau^{nK}{}_J
-\frac{1}{2}(\bar\phi^K_B\phi^A_L+\ep^{AC}\ep_{BD}\bar\phi^{K}_C\phi^D_{L})\phi^B_J\bar\phi^I_A
\tau_{m}{}^L{}_K\tau^{mJ}{}_I\bigg)
\ee
All equations of motions in (\ref{5dEOM1}) can be derived as Euler-Lagrange equations from the above action, and one can restore the continuous parameter $a_1$  by using (\ref{red}).

Using (\ref{vev})$-$(\ref{reality4}), one can reduce the law of supersymmetry transformations (\ref{susy5}) into
\e
\d\phi_m&=&-i\bar\ep^A\chi_{Am},\nonumber\\
\d\phi^A_I&=&-i\bar\ep^A\psi_I,
\nonumber\\
\d\chi_{Am}&=&\gamma^\a\ep_AD_\a\phi_m+\frac{i}{2}\gamma_{\a\b}\ep_AF^{\a\b}_m+i\ep_B
(\bar\phi^J_A\phi^B_I+\ep_{AC}\ep^{BD}\bar\phi^J_D\phi^C_I)\tau_m{}^I{}_J,
\nonumber\\
\d\psi_I&=&-2\gamma^\a\ep_A D_\a\phi^A_I-2i\ep_A\tau^{mJ}{}_I\phi_m\phi^A_J,\nonumber\\
\d A^m_\a&=&\bar\ep^A\gamma_\a\chi^m_{A}.\label{5dyms}
\ee
The action (\ref{action}) is invariant under the above supersymmetry transformations. If
(\ref{adjoint}) and (\ref{adjoint2}) are satisfied,
i.e., if the scalar fields $\phi^A_I$ and fermion fermionic fields $\psi_I$  are also in the adjoint representation of gauge group, we expect that the $\CN=1$ supersymmetry is enhanced to $\CN=2$, and theory is promoted to be the maximum supersymmetric Yang-Mills theory in 5D.

We now consider the possibility that $\langle C^\mu\rangle$ is a light-like vector. In Ref. \cite{LP, LS},
it was argued that if one uses the null reduction
\e
\langle C^\mu\rangle=g(1,0,\ldots,0,1),\quad\langle C^\mu \rangle\langle C_\mu \rangle=0,\label{null}
\ee
i.e. $\langle C^\mu\rangle$ is a light-like vector,
 the $(2,0)$ theory can be used
to describe a system of M5-branes.
So it is natural to expect that this $(1, 0)$ theory may be also used to describe multiple M5-branes  \cite{LR}. 
It would be interesting to explore this special $(1, 0)$ theory further. In particular, it would be interesting to introduce an additional abelian 3-form field into this $(1,0)$ theory (like Lambert and Sacco did in their work \cite{LS}), and see that whether the theory can be reduced to some 3D superconformal Chern-Simons matter theory or not.

Using the three equation of (\ref{eqs3}), one can solve $H^m_{\mu\nu\rho}$ in terms
of the field strength of the gauge field:
\e\label{hf}
C^2H^m_{\mu\nu\rho}=3F^m_{[\mu\nu}C_{\rho]}+\frac{1}{2}\vp_{\mu\nu\rho}{}^{\lambda\kappa\tau}F^m_{\lambda\kappa}C_\tau.
\ee
Substituting (\ref{null}) into (\ref{hf}), we find that the field strength obeys the duality condition:
\e\label{selfdual}
0=3F^m_{[\mu\nu}C_{\rho]}+\frac{1}{2}\vp_{\mu\nu\rho}{}^{\lambda\kappa\tau}F^m_{\lambda\kappa}C_\tau,
\ee
which can be decomposed into
\e
F^m_{\a 5}&=&F^m_{\a0},\\
F^m_{\a\b}&=&-\frac{1}{2}\vp_{\a\b}{}^{\g\d}F^m_{\g\d},\quad (\a, \b, \g, \d=1,\ldots,4.)
\ee
where $\vp_{1234}=\vp^{1234}=1$. We see that the field strength $F^m_{\a\b}$ is anti-selfdual.

Let $\langle C^\mu\rangle$ be a time-like vector, namely,
\e\label{time}
\langle C^\mu\rangle=g(1,0,\ldots,0).
\ee
Then the fields are covariantly static, that is
\e
0&=&D_0\phi_{m}=D_0 \phi_{m}=D_0\phi^A_I
=D_0\chi^A_{m}=D_0\psi_{I}=D_0 H_{\mu\nu\rho m}=\partial_0 g.
\ee
According to \cite{LP}, this theory may be a dual gauge theory for static 5-branes in 11 dimensional spacetime.

For more discussions on M5-branes and 6D $(1,0)$ and $(2,0)$ theories and 5D SYM theories, see \cite{5towers, 5kim, ABKS, 5Lambert, 5Lambert2, 5Douglas,5Hull,hhm,huang,hm}.



\section{Acknowledgement}
We are grateful to Jun-Bao Wu and Zhi-Guang Xiao for useful discussions. This work is supported in part by the National Science Foundation
of China (NSFC) under Grant No. 11475016, and supported partially by the Ren-Cai Foundation of Beijing Jiaotong University through Grant No. 2013RC029, and supported partially by the Scientific Research Foundation for Returned Scholars, Ministry of Education of China.

\appendix


\section{Closure of the $(1, 0)$ Tensor Multiplet Superalgebra}\label{Mini10}
In this section, we verify the closure of the poincare superalgebra of the $(1, 0)$ theory of Sec. \ref{secsmall}, using manifest $SU(2)$-notations. For convenience, we cite the supersymmetry transformations (\ref{susy3}) here
\e\label{susy9}
\d\phi_m&=&-i\bar\ep^A\chi_{Am},\nonumber\\
\d\chi_{Am}&=&\Gamma^\mu\ep_AD_\mu\phi_m+\frac{1}{3!}\frac{1}{2!}
\Gamma_{\mu\nu\lambda}\ep_AH^{\mu\nu\lambda}_m,
\nonumber\\
\d A_\mu^m&=&i\bar\ep^A\Gamma_{\mu\nu}\chi^m_{A}C^{\nu},
\nonumber\\
\d C^\nu&=&0,\nonumber\\
\d H_{\mu\nu\rho m}&=&
3i\bar\ep^{A}\Gamma_{[\mu\nu}D_{\rho]}\chi_{Am}-i\bar\ep^{A}\Gamma_{\mu\nu\rho\sigma}
C^{\sigma}\chi_{An}\phi_pf^{np}{}_{m},
\ee

The variation of the scalar fields reads
\e
[\d_1,\d_2]\phi_m=v^\mu D_\mu\phi_m,\label{scalar1}
\ee
where
\e
v^\mu\equiv-2i\bar\ep^A_2\Gamma^\mu\ep_{1A}.
\ee
It can be seen that the right-hand side of (\ref{scalar1}) is a covariant transformation.

Let us now consider the gauge fields. After some algebraic steps, one obtains
\e
[\d_1,\d_2]A^m_\mu&=&v^\nu F^m_{\nu\mu}-D_\mu\Lambda^m\nonumber\\
&&+v^\nu(F^m_{\mu\nu}-H^m_{\mu\nu\rho}C^\rho)\nonumber\\
&&-v^\mu C^\nu D_\nu\phi^m,\label{gauge2}
\ee
where
\e\label{Lambda}
\Lambda^m\equiv-v^\nu C_\nu\phi^m.
\ee
The first term of the first line of (\ref{gauge2}) is the covariant translation, while the second term is a  gauge transformation. The second line and third line must be the equations of motion:
\e
0&=&F^m_{\mu\nu}-H^m_{\mu\nu\rho}C^\rho,\label{gauge3}\\
0&=&C^\nu D_\nu\phi^m.\label{yphi}
\ee
A super-variation on $0=C^\nu D_\nu\phi^m$ gives
\e
0=C^\nu D_\nu\chi^m_A.\label{ychi}
\ee


By the definition of $\Lambda$ (see (\ref{Lambda})), we see that $[\Lambda,\phi]=0$. So equation (\ref{scalar1}) can be recast into the expected form
\e
[\d_1,\d_2]\phi_m=v^\mu D_\mu\phi_m+[\Lambda,\phi]_m.
\ee

We now check the closure on the fermionic fields. A lengthy calculation gives
\e\label{fermion1}
[\d_1,\d_2]\chi_{Am}&=&v^\mu D_\mu\chi_{Am}+[\Lambda,\chi_A]_m\nonumber\\
&&-\frac{1}{4}v^\mu\Gamma_\mu(\Gamma_\nu D^\nu\chi_{Am}+\Gamma_\nu C^\nu[\phi,\chi_A]_m)
\ee
Clearly, the second line must be the equations of motion for the fermions:
\e
0=\Gamma_\nu D^\nu\chi_{Am}+\Gamma_\nu C^\nu[\phi,\chi_A]_m.\label{eomfermi}
\ee
In computing (\ref{fermion1}), we have used the Fierz identity (\ref{Fierz1}).
As observed in \cite{LP}, the equations of motion of the fermions (\ref{eomfermi}) can be also derived by requiring $\d H_{\mu\nu\rho m}$ to obey the self-dual conditions
\e
\d H_{\mu\nu\rho m}=\frac{1}{3!}\vp_{\mu\nu\rho\s\lambda\tau}\d H^{\s\lambda\tau}.
\ee

As for the auxiliary field $C^\mu$, we have $[\d_1,\d_2]C^\mu=0$. On the other hand, we expect
\e
[\d_1,\d_2]C^\mu&=&v^\nu D_\nu C^\mu + [\Lambda, C^\mu]
\ee
However, since $C^\mu$ is not ``charged" by the gauge group, we must have $[\Lambda, C^\mu]=0$, leading to
\e
D_\nu C^\mu=\partial_\nu C^\mu=0,\label{yc}
\ee
i.e. $C^\mu$ is a constant field.

Finally, we compute the super-variations of the tensor fields:
\e
&&[\d_1,\d_2]H_{\mu\nu\rho m}\nonumber
\\&=& v^\s D_\s H_{\mu\nu\rho m}+[\Lambda, H_{\mu\nu\rho}]_m\nonumber\\
&&+3(v_{[\mu}[F_{\nu\rho]}, \phi]_m-v_{[\mu}[H_{\nu\rho]\lambda}C^\lambda,\phi]_m)
\\
&&+
4v^\s\bigg(D_{[\mu}H_{\nu\rho\s]m}+\frac{i}{8}\vp_{\mu\nu\rho\lambda\s\tau}
(\bar\chi^A_{n}\Gamma^\tau
\chi_{Ap})C^\lambda f^{np}{}_m+\frac{1}{4}\vp_{\mu\nu\rho\lambda\s\tau}\phi_nC^\lambda D^\tau\phi_p f^{np}{}_m\bigg).\nonumber
\ee
The second line vanishes by equation (\ref{gauge3}); the third line turns out to be the equations of motions for the tensor fields:
\e
0=D_{[\mu}H_{\nu\rho\s]m}+\frac{i}{8}\vp_{\mu\nu\rho\lambda\s\tau}
(\bar\chi^A_{n}\Gamma^\tau
\chi_{Ap})C^\lambda f^{np}{}_m+\frac{1}{4}\vp_{\mu\nu\rho\lambda\s\tau}\phi_nC^\lambda D^\tau\phi_p f^{np}{}_m.\label{eomtensor1}
\ee

Combining the Bianchi identity $D_{[\mu}F^m_{\nu\rho]}=0$ and the equations of motion $F^m_{\nu\rho}=H^m_{\nu\rho\lambda}C^\lambda$ (see (\ref{gauge3})), we learn that
$D_{[\mu}H^m_{\nu\rho]\lambda}C^\lambda=0$, which is equivalent to
\e
\frac{4}{3}C^\lambda D_{[\mu}H^m_{\nu\rho\lambda]}+\frac{1}{3}C^\lambda D_\lambda H^m_{\mu\nu\rho}=0.
\ee
However, the first term vanishes by the equations of motion (\ref{eomtensor1}). We therefore  have the constraint equation:
\e
C^\lambda D_\lambda H^m_{\mu\nu\rho}=0.\label{ytensor}
\ee
The above equation implies that
\e
C^\lambda D_\lambda F^m_{\mu\nu}=0,\label{ygauge}
\ee
which can be also derived by using the Bianchi identity $D_{[\mu}F^m_{\nu\rho]}=0$ and the equations of motion $F^m_{\nu\rho}=H^m_{\nu\rho\lambda}C^\lambda$ (see (\ref{gauge3})).

Taking a super-variation on the equations of motion for the fermions
\e
\d(\Gamma^\mu D_\mu\chi^A_m-\Gamma^\mu C_\mu[\chi^A,\phi]_m)=0,
\ee
one obtains the equations of motion (\ref{gauge3}) and (\ref{eomtensor1}), and the equations of motion for the scalar fields:
\e
0=D^2\phi_m-\frac{i}{2}(\bar\chi^A_n\Gamma_\mu\chi_{Ap})C^\mu f^{np}{}_m.\label{eomscalar}
\ee

In summary, the equations of motion (\ref{gauge3}), (\ref{eomfermi}), (\ref{eomtensor1}), and (\ref{eomscalar}) are in agreement with (\ref{eqs1}); And the constraint equations (\ref{yphi}), (\ref{ychi}), (\ref{yc}), (\ref{ytensor}), and (\ref{ygauge}) are exactly the same as the last line of (\ref{eqs1}).

\section{Super-variations of the Equations of Motion of the $(1,0)$ Theory}\label{secsveom}

In this section, we will check that the set of equations of motion (\ref{eqs3}) in section \ref{sec1scft} are closed under supersymmetry transformations (\ref{susy5}).

First of all, we have already learned that the super-variation
\e
0&=&\d(C^\s D_\s\phi^{m})
\ee
gives
\e
0=C^\s D_\s\chi^{m}_A.
\ee
(See also (\ref{yphi}) and (\ref{ychi}).) Taking a super-variation on the above equation,
\e
0=\d(C^\rho D_\rho\chi^{m}_A),
\ee
one obtains
\e
0&=&\Gamma^\mu\ep_A[[C^\rho F_{\rho\mu},\phi]_m+D_\mu(C^\rho D_\rho\phi_m)]\nonumber\\&&+\frac{1}{12}\Gamma_{\mu\nu\lambda}\ep_A C^\rho D_\rho H^{\mu\nu\lambda}_m\nonumber\\&&+
\Gamma_\lambda\ep_BC^\lambda [C^\rho D_\rho(\bar\phi^J_A\phi^B_I+\bar\phi^{BJ}\phi_{AI})]\tau_m^I{}_J.
\ee
By $F_{\rho\mu}=H_{\rho\mu\nu}C^\nu$ (see the third equation of (\ref{eqs3})), the first term of the first line vanishes; The rest terms are the following constraint equations
\e
0=C^\rho D_\rho\phi_m=C^\rho D_\rho H^{\mu\nu\lambda}_m=C^\rho D_\rho\bar\phi^J_A=C^\rho D_\rho\phi^B_I.
\ee

As for the self-dual field strength, we have
\e\label{dcH}
0&=&\d(C^\lambda D_\lambda H_{\mu\nu\rho m})\nonumber\\
&=&C^\lambda[i\bar\ep^A\Gamma_{\lambda\s}\chi_AC^\s, H_{\mu\nu\rho}]_m\nonumber\\
&&+C^\lambda D_\lambda(3i\bar\ep^{A}\Gamma_{[\mu\nu}D_{\rho]}\chi_{Am}-i\bar\ep^{A}\Gamma_{\mu\nu\rho\sigma}
C^{\sigma}\chi_{An}\phi_pf^{np}{}_{m}\nonumber\\
&&-i\bar\ep^A\Gamma_{\mu\nu\rho\s}\psi_IC^\s
\bar\phi^J_A\tau_m{}^I{}_J+i\bar\psi^I\Gamma_{\mu\nu\rho\s}\ep_A\phi^A_J\tau_m{}^J{}_I)
\ee
Since $\Gamma_{\lambda\s}C^\lambda C^\s=0$, the first line of (\ref{dcH}) vanishes; Using $0=C^\lambda D_\lambda\chi_{Am}=C^\lambda F_{\lambda\rho}$, the first term of the second line vanishes; Using $0=C^\lambda D_\lambda\chi_{Am}=C^\lambda D_\lambda\phi_p$, the second term of the second line vanishes; The terms of the third line are the constraint equations:
\e
0=C^\lambda D_\lambda\bar\phi^J_A=C^\lambda D_\lambda\phi^A_J=C^\lambda D_\lambda\psi_I=C^\lambda D_\lambda\bar\psi^I
\ee

We now calculate the super-variation of the constraint equation for the scalar fields $\phi^A_I$:
\e
0&=&\d(C^\lambda D_\lambda\phi^A_I)\nonumber\\
&=& C^\lambda(i\bar\ep^A\Gamma_{\lambda\nu}\chi^m_{A}C^{\nu})(-\tau_m)^J{}_I\phi^A_J
\nonumber\\
&&+C^\lambda D_\lambda(i\bar\ep^A\psi_I)\label{dphii}
\ee
Since $\Gamma_{\lambda\nu}C^\lambda C^{\nu}=0$, the first line of (\ref{dphii}) vanishes; The second line is nothing but
$
0=C^\lambda D_\lambda\psi_I.
$
By the reality condition $\bar\phi_A^I=(\phi^A_I)^*$, the equation $0=\d(C^\lambda D_\lambda\bar\phi_A^I)$ must be also satisfied.

We now turn to the constraint equation for fermionic fields $\psi_I$,
\e
0&=&\d(C^\lambda D_\lambda\psi_I)\nonumber\\
&=& C^\lambda(i\bar\ep^A\Gamma_{\lambda\nu}\chi^m_{A}C^{\nu})(-\tau_m)^J{}_I\psi_J
\nonumber\\
&&+C^\lambda D_\lambda(-2\ep_A\Gamma^\mu D_\mu \phi^A_I-2\Gamma_\rho\ep_AC^\rho\tau^{mJ}{}_I\phi_m\phi^A_J)\label{dpsij}
\ee
The first line of (\ref{dpsij}) vanishes due to that $\Gamma_{\lambda\nu}C^\lambda C^{\nu}=0$; By $0=C^\lambda D_\lambda\phi^A_I=C^\lambda F_{\lambda\mu}=C^\lambda D_\lambda\phi_m$, we see that the second line of (\ref{dpsij}) also vanishes. Because of the reality condition $\psi^I=(\psi_I)^*$, the equation $0=\d(C^\lambda D_\lambda\psi^I)$ must also hold.

Let us consider the super-variation of the equations of motion for the fermionic fields $\chi_{Am}$:
\e
0=\d(\Gamma^\mu D_\mu\chi_{Am}+\Gamma^\mu C_\mu[\phi, \chi_A]_m+\Gamma^\mu C_\mu B^{-1}\psi^{J}\phi_{AI}\tau_{m}{}^I{}_J
-\Gamma^\mu C_\mu\psi_{J}\bar\phi^{I}_A\tau_{m}{}^J{}_I)
\ee
A straightforward calculation gives
\e
0&=&\ep_A\bigg(D^2\phi_m-\frac{i}{2}C^\nu(\bar\chi^B_{n}\Gamma_\nu\chi_{Bp})f^{np}{}_m
+\frac{i}{2}C^\nu(\bar\psi^J\Gamma_\nu\psi_I)\tau_m{}^I{}_J+2C^2
\bar\phi^J_A\phi^A_I\phi^n(\tau_{(m}\tau_{n)})^I{}_J\bigg)
\nonumber\\
&&+
\frac{1}{6}\Gamma^{\mu\nu\rho\s}\ep_A\bigg[D_{[\mu}H_{\nu\rho\s]m}-\frac{i}{8}\vp_{\mu\nu\rho\s\lambda\tau}C^\lambda \bigg((\bar\chi^A_n\Gamma^\tau
\chi_{Ap}) f^{np}{}_m+(\bar\psi^J\Gamma^\tau\psi_I)\tau_m{}^I{}_J\bigg)\nonumber\\
&&\quad\quad\quad-\frac{1}{4}\vp_{\mu\nu\rho\s\lambda\tau}C^\lambda\bigg(\phi_nD^\tau\phi_p f^{np}{}_m+\bar\phi^J_BD^\tau\phi^B_I\tau_m{}^I{}_J
-\phi^B_JD^\tau\bar\phi^I_B
\tau_m^J{}_I\bigg)\bigg]
\nonumber\\
&&+\frac{1}{2}\Gamma^{\mu\nu}\ep_A[F_{\mu\nu}-H_{\mu\nu\rho}C^\rho,\phi]_m.\label{dpsim}
\ee
It can be seen that the second and third lines are the equations of motion for the tensor fields, while the last line is the equations of motion for the gauge fields. So the first line must be the equations of motion for the scalar fields $\phi_m$. In deriving (\ref{dpsim}), we have used the constraint equations $0=C^\mu D_\mu\phi_m=C^\mu D_\mu\bar\phi^I_A=C^\mu D_\mu\phi_{AI}$.

We now study the supersymmetry transformation of the equations of motion for the fermionic fields $\psi_I$:
\e
0&=&\d(\Gamma^\mu D_\mu\psi_I+\Gamma^\mu C_\mu\tau^{mJ}{}_I\phi_m\psi_J-2\Gamma^\mu C_\mu\psi_{Am}\tau^{mJ}{}_I\phi^A_J)
\ee
The result is
\e
0&=&\Gamma^{\mu\nu}\ep_{A}(F^m_{\mu\nu}-H^m_{\mu\nu\rho}C^\rho)\phi^A_J\tau^{mJ}{}_I\nonumber\\
&&-2\ep_{A}[D^2\phi^A_I-iC^\nu(\bar\chi^{A}_m\Gamma_\nu\psi_J)\tau^{mJ}{}_I
+C^2\phi^A_K\phi_m\phi_n\tau^{mK}{}_J\tau^{nJ}{}_I\nonumber\\
&&+C^2(\bar\phi^K_B\phi^A_L+\bar\phi^{AK}\phi_{BL})\phi^B_J\tau_{m}{}^L{}_K\tau^{mJ}{}_I
],\label{dpsia}
\ee
The first line is the equations of motion for the gauge fields. So the second and third lines must be the equations of motion for scalar fields $\phi^A_I$. In deriving (\ref{dpsia}), we have also used  $0=C^\mu D_\mu\phi_m=C^\mu D_\mu\bar\phi^I_A=C^\mu D_\mu\phi_{AI}$.

The super-variation of the equations of motion for the gauge fields is given by
\e
0=\d(F^m_{\mu\nu}-H^m_{\mu\nu\rho}C^\rho),
\ee
which is equivalent to
\e
0=-i\bar\ep^A\Gamma_{\mu\nu}(C^\rho D_\rho\chi^m_{A}),
\ee
i.e. the constraint equations for the fermionic fields $\chi^m_A$.

Under supersymmetry transformations (\ref{susy5}), the 6th equation of (\ref{eqs3}) becomes
\e\label{dhg}
0&=&\d\bigg[D_{[\mu}H_{\nu\rho\s]m}-\frac{i}{8}\vp_{\mu\nu\rho\s\lambda\tau}C^\lambda \bigg((\bar\chi^A_n\Gamma^\tau
\chi_{Ap}) f^{np}{}_m+(\bar\psi^J\Gamma^\tau\psi_I)\tau_m{}^I{}_J\bigg)\nonumber\\
&&\quad\quad\quad-\frac{1}{4}\vp_{\mu\nu\rho\s\lambda\tau}C^\lambda\bigg(\phi_nD^\tau\phi_p f^{np}{}_m+\bar\phi^J_BD^\tau\phi^B_I\tau_m{}^I{}_J
-\phi^B_JD^\tau\bar\phi^I_B
\tau_m^J{}_I\bigg)\bigg]
\ee
The above equation is complicated, hence it is not easy to verify it directly. Our strategy is to take care of a simpler version of (\ref{dhg}) first: \emph{Without} coupling to the matter fields, the equations of motion of $H_{\mu\nu\rho}$ are given by fourth equation of (\ref{eqs2}). Under the supersymmetry transformations (\ref{susy3}), the fourth equation of (\ref{eqs2}) should obey\footnote{To distinguish the supersymmetry transformations (\ref{susy3}) and (\ref{susy5}), in this appendix, we replace the super-variation ``$\d$" in (\ref{susy3}) by ``$\bar\d$", while the super-variation  in (\ref{susy5}) is still denoted as ``$\d$".}
\begin{equation}
0=\bar\d\bigg(D_{[\mu}H_{\nu\rho\s]p}+\frac{i}{8}\vp_{\mu\nu\rho\lambda\s\tau}
(\bar\chi^A_{m}\Gamma^\tau
\chi_{An})C^\lambda f^{mn}{}_p+\frac{1}{4}\vp_{\mu\nu\rho\lambda\s\tau}\phi_mD^\tau\phi_nC^\lambda f^{mn}{}_p\bigg)\label{dhs}
\end{equation}
After verifying the above equation, it will be much easier to verify (\ref{dhg}), since the proof of the above equation can be used to verify (\ref{dhg}). Under (\ref{susy3}), Eq. (\ref{dhs}) reads
\e
0&=&D_{[\mu}\bar\d H_{\nu\rho\s]m}+[\bar\d A_{[\mu},H_{\nu\rho\s]}]_m\nonumber\\
&&-\frac{1}{4}\vp_{\mu\nu\rho\s\lambda\tau}C^\tau([D^\lambda\bar\d\phi,\phi]_m+
[[\bar\d A^\lambda,\phi],\phi]_m+[D^\lambda\phi,\bar\d\phi]_m)\nonumber\\
&&-\frac{i}{4}\vp_{\mu\nu\rho\s\lambda\tau}C^\lambda(\bar\chi^A_n\Gamma^\tau\bar\d\chi_{Ap})f^{np}{}_m
\nonumber\\
&=&\frac{3i}{2}\bar\ep^A\Gamma_{[\nu\rho}[F_{\mu\s]},\chi_A]_m+i\bar\ep^A\Gamma_{\lambda[\nu
\rho\s}[D_{\mu]}\chi_A,\phi]_mC^\lambda-i\bar\ep^A\Gamma_{\lambda[\nu
\rho\s}[D_{\mu]}\phi,\chi_A]_mC^\lambda\nonumber\\
&&
-i\bar\ep^A\Gamma_{\lambda[\mu}H_{\nu\rho\s]p}\chi_{An}C^\lambda f^{np}{}_m
\nonumber\\
&&+\frac{i}{4}\vp_{\mu\nu\rho\s\lambda\tau}C^\tau\bigg([\bar\ep^AD^\lambda\chi_A,\phi]_m
-C_\kappa[[\bar\ep^A\Gamma^{\lambda\kappa}\chi_A,\phi],\phi]_m
+[D^\lambda\phi,\bar\ep^A\chi_A]_m\bigg)
\label{dhs2}\\
&&-\frac{i}{4}\vp_{\mu\nu\rho\s\lambda\tau}C^\lambda\bigg((\bar\chi^A_n\Gamma^{\tau\kappa}\ep_A)
D_\kappa\phi_p+(\bar\chi^A_n\ep_A)
D^\tau\phi_p+\frac{1}{12}(\bar\chi^A_n\Gamma^{\tau}\Gamma^{\kappa\xi\delta
}\ep_A)H_{\kappa\xi\d}
\bigg)f^{np}{}_m\nonumber
\ee
Note that the third term of the third line of (\ref{dhs2}) cancels the second term of the fourth line; We group the rest terms of (\ref{dhs2}) as follows
\e
0&=&\bigg(\frac{3i}{2}\bar\ep^A\Gamma_{[\nu\rho}[F_{\mu\s]},\chi_A]_m-i\bar\ep^A\Gamma_{\lambda[\mu}H_{\nu\rho\s]p}\chi_{An}C^\lambda f^{np}{}_m-\frac{i}{48}\vp_{\mu\nu\rho\s\lambda\tau}C^\lambda(\bar\chi^A_n\Gamma^{\tau}\Gamma^{\kappa\xi\delta
}\ep_A)H_{\kappa\xi\d}
f^{np}{}_m\bigg)\nonumber\\
&&+\bigg[i\bar\ep^A\Gamma_{\lambda[\nu
\rho\s}[D_{\mu]}\chi_A,\phi]_mC^\lambda+\frac{i}{4}\vp_{\mu\nu\rho\s\lambda\tau}C^\tau\bigg([\bar\ep^AD^\lambda\chi_A,\phi]_m
-C_\kappa[[\bar\ep^A\Gamma^{\lambda\kappa}\chi_A,\phi],\phi]_m\bigg)\bigg]
\nonumber\\
&&-\bigg(
i\bar\ep^A\Gamma_{\lambda[\nu
\rho\s}[D_{\mu]}\phi,\chi_A]_mC^\lambda+\frac{i}{4}\vp_{\mu\nu\rho\s\lambda\tau}C^\lambda(\bar\chi^A_n\Gamma^{\tau\kappa}\ep_A)
D_\kappa\phi_p\bigg)\label{dhs3}
\ee

Using the self-dual conditions (\ref{selfdual0}), the last term of the first line of (\ref{dhs3})
can be rewritten as
\e
-\frac{3i}{2}\bar\ep^A\Gamma_{[\nu\rho}[F_{\mu\s]},\chi_A]_m+i\bar\ep^AC^\lambda
\Gamma_{\lambda[\mu}H_{\nu\rho\s]p}\chi_{An}f^{np}{}_m.
\ee
The above two terms cancel the first term of the first line of (\ref{dhs3}), so the first line of (\ref{dhs3}) vanishes.

Using $C^\lambda D_\lambda\chi_{Am}=0$, the first term of the second line of (\ref{dhs3}) can be written as
\e
\frac{5i}{4}\bar\ep^A\Gamma_{[\lambda\nu
\rho\s}[D_{\mu]}\chi_A,\phi]_mC^\lambda.
\ee
So the second line of (\ref{dhs3}) becomes
\begin{equation}\label{dhs4p}
\bigg[\frac{5i}{4}\bar\ep^A\Gamma_{[\lambda\nu
\rho\s}[D_{\mu]}\chi_A,\phi]_mC^\lambda+\frac{i}{4}\vp_{\mu\nu\rho\s\lambda\tau}C^\tau\bigg([\bar\ep^AD^\lambda\chi_A,\phi]_m
-C_\kappa[[\bar\ep^A\Gamma^{\lambda\kappa}\chi_A,\phi],\phi]_m\bigg)\bigg].
\end{equation}
The above expression is zero by the equations of motion for the fermions $\chi_{Am}$ (see the third equation of (\ref{eqs2})). To see this, we multiply the third equation of (\ref{eqs2}) by $\Gamma^\nu$,
\e
0&=&\Gamma^\nu(\Gamma^\mu D_\mu\chi_{Am}-\Gamma^\mu C_\mu\chi_{An}\phi_pf^{np}{}_m),
\ee
which is
\e
0=\Gamma^{\nu\mu}D_\mu\chi_{Am}+D^\nu\chi_{Am}+\Gamma^{\nu\mu}C_\mu[\phi,\chi_A]_m+C^\nu[\phi,\chi_A]_m;\label{dhs4}
\ee
Relabeling the indices ($\nu\rightarrow\lambda$ and $\mu\rightarrow\kappa$), and multiplying both sides by $\frac{i}{4}\vp_{\mu\nu\rho\s\lambda\tau}C^\tau$, Eq. (\ref{dhs4}) becomes
\e\label{dhs5}
0=\frac{5i}{4}\Gamma_{[\lambda\nu
\rho\s}D_{\mu]}\chi_{Am}C^\lambda+\frac{i}{4}\vp_{\mu\nu\rho\s\lambda\tau}C^\tau\bigg(D^\lambda\chi_{Am}
-C_\kappa[\Gamma^{\lambda\kappa}\chi_A,\phi]_m\bigg).
\ee
Multiplying (\ref{dhs5}) by $\bar\ep^A$, and then taking commutator with $\phi$, the right hand side turns out to be exactly the same as (\ref{dhs4p}), so (\ref{dhs4p}) must vanish.

Using $C^\lambda D_\lambda\phi_{p}=0$, one can show that the last line of (\ref{dhs3}) also vanishes. This finishes the proof of (\ref{dhs}).

We are ready to verify (\ref{dhg}). We begin by proving three important equations which are useful in verifying (\ref{dhg}). In exactly the same way for deriving (\ref{dhs5}), we multiply the fifth equation  of (\ref{eqs3}) by $\frac{i}{4}\vp_{\mu\nu\rho\s\lambda\tau}C^\tau\Gamma^\nu$; The result is
\e
0&=&\frac{5i}{4}\Gamma_{[\lambda\nu
\rho\s}D_{\mu]}\chi_{Am}C^\lambda+\frac{i}{4}\vp_{\mu\nu\rho\s\lambda\tau}C^\tau\bigg(D^\lambda\chi_{Am}
-C_\kappa[\Gamma^{\lambda\kappa}\chi_A,\phi]_m\bigg)\nonumber\\
&&+\frac{i}{4}\vp_{\mu\nu\rho\s\lambda\tau}C^\tau C_\eta\Gamma^{\lambda\eta}(B^{-1}\psi^{ J\dag}\phi_{AI}-\psi_I\bar\phi^J_A)\tau_m^I{}_J.\label{dhs8}
\ee
As a check, if we set $\psi_I=0$, then (\ref{dhs8}) is reduced to (\ref{dhs5}). Multiplying (\ref{dhs8}) by $\bar\ep^A$, and then taking commutator with $\phi$, we obtain
\e
0&=&\frac{5i}{4}\bar\ep^A\Gamma_{[\lambda\nu
\rho\s}[D_{\mu]}\chi_A,\phi]_mC^\lambda+\frac{i}{4}\vp_{\mu\nu\rho\s\lambda\tau}C^\tau\bigg([\bar\ep^AD^\lambda\chi_A,\phi]_m
-C_\kappa[[\bar\ep^A\Gamma^{\lambda\kappa}\chi_A,\phi],\phi]_m\bigg)\nonumber\\
&&+\frac{i}{4}\vp_{\mu\nu\rho\s\lambda\tau}C^\tau C_\eta\bar\ep^A\Gamma^{\lambda\eta}(B^{-1}\psi^{ J\dag}\phi_{AI}-\psi_I\bar\phi^J_A)\phi_pf^{np}{}_m\tau_n^I{}_J.\label{dhs9}
\ee
Notice that the first line of (\ref{dhs9}) is exactly the same as (\ref{dhs4p}). This is expected, since now the tensor multiplets are coupling with the hypermultiplets.

Similarly, multiplying the EOM of $\psi_I$  or  the fourth equation  of (\ref{eqs3})
\e
0&=&\Gamma^\eta D_\eta\psi_I+\Gamma^\eta C_\eta\tau^{nK}{}_I\phi_n\psi_K-2\Gamma^\eta C_\eta\chi_{An}\tau^{nK}{}_I\phi^A_K,
\ee
by $(C^\tau\bar\phi^J_A\tau_m^I{}_J)\frac{i}{4}\vp_{\mu\nu\rho\s\lambda\tau}\bar\ep^A\Gamma^\lambda$,
we are able to derive
\e
0&=&\frac{i}{4}\vp_{\mu\nu\rho\s\lambda\tau}\bigg((\bar\ep^A\Gamma^\lambda\Gamma^\eta D_\eta\psi_I)C^\tau\bar\phi^J_A\tau_m^I{}_J
-(\bar\ep^A\Gamma^{\tau\eta}\psi_J)C^\lambda C_\eta\phi^n\bar\phi^K_A\tau_n^J{}_I\tau_m^I{}_K\label{dhs10}\\
&&-(\bar\ep^A\Gamma^{\lambda\eta}\chi_A^n)C^\tau C_\eta\phi^B_J\bar\phi^K_B\tau_n^J{}_I\tau_m^I{}_K+(\bar\chi^{Bn}\Gamma^{\tau\eta}\ep_A)C^\lambda C_\eta(\phi_{BJ}\bar\phi^{AK}+\phi^A_J\bar\phi^K_B)\tau_n^J{}_I\tau_m^I{}_K\bigg)\nonumber
\ee
The conjugate equation of (\ref{dhs10}) is
\e
0&=&\frac{i}{4}\vp_{\mu\nu\rho\s\lambda\tau}\bigg((\overline{D_\eta\psi^I}\Gamma^\eta \Gamma^\lambda \ep_A)C^\tau\bar\phi_J^A\tau_m^J{}_I
-(\bar\psi^J\Gamma^{\tau\eta}\ep_A)C^\lambda C_\eta\phi^n\phi_K^A\tau_n^I{}_J\tau_m^K{}_I\label{dhs11}\\
&&+(\bar\ep^A\Gamma^{\lambda\eta}\chi_A^n)C^\tau C_\eta\bar\phi_B^J\phi_K^B\tau_n^I{}_J\tau_m^K{}_I+(\bar\chi^{Bn}\Gamma^{\tau\eta}\ep_A)C^\lambda C_\eta(\phi_{BK}\bar\phi^{AJ}+\phi^A_K\bar\phi^J_B)\tau_n^I{}_J\tau_m^K{}_I\bigg)\nonumber
\ee

We now try to calculate (\ref{dhg}). Taking account of the relation of (\ref{susy5}) and (\ref{susy3}), we find that under supersymmetry transformations (\ref{susy5}),
(\ref{dhg}) becomes
\e
0&=&D_{[\mu}\bar\d H_{\nu\rho\s]m}+[\bar\d A_{[\mu},H_{\nu\rho\s]}]_m\nonumber\\
&&-\frac{1}{4}\vp_{\mu\nu\rho\s\lambda\tau}C^\tau([D^\lambda\bar\d\phi,\phi]_m+
[[\bar\d A^\lambda,\phi],\phi]_m+[D^\lambda\phi,\bar\d\phi]_m)\nonumber\\
&&-\frac{i}{4}\vp_{\mu\nu\rho\s\lambda\tau}C^\lambda(\bar\chi^A_n\Gamma^\tau\bar\d\chi_{Ap})f^{np}{}_m
\nonumber\\
&&-i\bar\ep^AD_{[\mu}(\Gamma_{\nu\rho\s]\lambda}C^\lambda\psi_I\tau_m^I{}_J\bar\phi^J_A)\nonumber\\
&&+iD_{[\mu}(\bar\psi^I\Gamma_{\nu\rho\s]\lambda}C^\lambda\ep_A(\tau_m)^J{}_I\phi_J^A)\nonumber\\
&&+\frac{1}{4}\vp_{\mu\nu\rho\s\lambda\tau}C^\tau\bigg(\d\bar\phi^J_BD^\lambda\phi^B_I
+\bar\phi^J_B\d A^\lambda_n(-\tau^{nK}{}_I)\phi^B_K+\bar\phi^J_BD^\lambda\d\phi^B_I\nonumber\\
&&-\d\phi^B_I D_\lambda\bar\phi^J_B-\phi_I^B\d A_\lambda^n(\tau_{n}){}^J{}_K\bar\phi^K_B-\phi_I^BD^\lambda\d\bar\phi_B^J\bigg)(\tau_{m}){}^I{}_J\nonumber\\
&&-\frac{i}{8}\vp_{\mu\nu\rho\s\lambda\tau}C^\lambda(\d'\bar\chi^A_n)\Gamma^\tau\chi_{Ap}f^{np}{}_m\nonumber\\
&&-\frac{i}{8}\vp_{\mu\nu\rho\s\lambda\tau}C^\lambda\bar\chi^A_n\Gamma^\tau(\d'\chi_{Ap})f^{np}{}_m\nonumber\\
&&-\frac{i}{8}\vp_{\mu\nu\rho\s\lambda\tau}C^\lambda(\d\bar\psi^J)\Gamma^\tau\psi_I\tau_m^I{}_J\nonumber\\
&&-\frac{i}{8}\vp_{\mu\nu\rho\s\lambda\tau} C^\lambda\bar\psi^J\Gamma^\tau(\d\psi_I)\tau_m^I{}_J,\label{dhs6}
\ee
where ``$\bar\d$" and ``$\d$" refer to the super-variations in (\ref{susy3}) and (\ref{susy5}), respectively, and
\e
\d'\chi_{Ap}\equiv\Gamma_\eta\ep_BC^\eta(\bar\phi^K_A\phi^B_L+\bar\phi^{BK}\phi_{AL})\tau_p^L{}_K.\label{dprime}
\ee
Notice that the first three lines of (\ref{dhs5}) are nothing but (\ref{dhs2}).
Using the results for proving (\ref{dhs2}) and (\ref{dhs3}), and using Eq. (\ref{dhs9}), the first three lines of
(\ref{dhs6}) turn out to be
\e
\frac{i}{4}\vp_{\mu\nu\rho\s\lambda\tau}C^\tau C_\eta\bar\ep^A\Gamma^{\lambda\eta}(B^{-1}\psi^{ J\dag}\phi_{AI}-\psi_I\bar\phi^J_A)\phi_pf^{np}{}_m\tau_n^I{}_J\label{thcoumpling}
\ee
Plugging (\ref{thcoumpling}) into (\ref{dhs6}), and using (\ref{susy5}) and (\ref{dprime}), we obtain
\e
0&=&\frac{i}{4}\vp_{\mu\nu\rho\s\lambda\tau}C^\tau C_\eta\bar\ep^A\Gamma^{\lambda\eta}(B^{-1}\psi^{ J\dag}\phi_{AI}-\psi_I\bar\phi^J_A)\phi_pf^{np}{}_m\tau_n^I{}_J\nonumber\\
&&+\frac{i}{4}\vp_{\mu\nu\rho\s\lambda\tau}(\bar\ep^A\Gamma^{\lambda\eta} D_\eta\psi_I)C^\tau\bar\phi^J_A\tau_m^I{}_J \nonumber\\
&&+\frac{i}{4}\vp_{\mu\nu\rho\s\lambda\tau}(\overline{D_\eta\psi^I}\Gamma^{\lambda \eta} \ep_A)C^\tau\bar\phi_J^A\tau_m^J{}_I\nonumber\\
&&-\frac{i}{4}\vp_{\mu\nu\rho\s\lambda\tau}(\bar\ep^A\Gamma^{\lambda\eta}\chi_A^n)C^\tau C_\eta\phi^B_J\bar\phi^K_B(\tau_n^J{}_I\tau_m^I{}_K+\tau_m^J{}_I\tau_n^I{}_K)\nonumber\\
&&+\frac{i}{4}\vp_{\mu\nu\rho\s\lambda\tau}(\bar\ep^A D_\lambda\psi_I)C^\tau\bar\phi^J_A\tau_m^I{}_J\nonumber\\
&&-\frac{i}{4}\vp_{\mu\nu\rho\s\lambda\tau}(\overline{D^\lambda\psi^I} \ep_A)C^\tau\phi_J^A\tau_m^J{}_I\nonumber\\
&&-\frac{i}{4}\vp_{\mu\nu\rho\s\lambda\tau}(\bar\chi^{A}_n\Gamma^{\tau\eta}\ep_B)C^\lambda C_\eta(\bar\phi^K_{A}\phi^B_{L}+\bar\phi^{BK}\phi_{AL})\tau_p^L{}_Kf^{np}{}_m\nonumber\\
&&-\frac{i}{4}\vp_{\mu\nu\rho\s\lambda\tau}(\bar\ep^A\Gamma^{\tau\eta}\psi_J)C^\lambda C_\eta\phi^n\bar\phi^K_A\tau_m^J{}_I\tau_n^I{}_K\nonumber\\
&&+\frac{i}{4}\vp_{\mu\nu\rho\s\lambda\tau} (\bar\psi^J\Gamma^{\tau\eta}\ep_A)C^\lambda C_\eta\phi^n\phi_K^A\tau_n^K{}_I\tau_m^I{}_J\label{dhs7}
\ee
Substituting (\ref{dhs10}) and (\ref{dhs11}) into (\ref{dhs7}), one obtains
\e
0&=&\frac{i}{4}\vp_{\mu\nu\rho\s\lambda\tau}C^\tau C_\eta\bar\ep^A\Gamma^{\lambda\eta}(B^{-1}\psi^{ J\dag}\phi_{AI}-\psi_I\bar\phi^J_A)\phi_pf^{np}{}_m\tau_n^I{}_J\nonumber\\
&&-\frac{i}{4}\vp_{\mu\nu\rho\s\lambda\tau}(\bar\ep^A\Gamma^{\tau\eta}\psi_J)C^\lambda C_\eta\phi^n\bar\phi^K_A(\tau_m^J{}_I\tau_n^I{}_K-\tau_n^J{}_I\tau_m^I{}_K)
\nonumber\\
&&-\frac{i}{4}\vp_{\mu\nu\rho\s\lambda\tau}(\bar\psi^J\Gamma^{\tau\eta}\ep_A)C^\lambda C_\eta\phi^n\bar\phi^A_K(\tau_n^K{}_I\tau_m^I{}_J-\tau_m^K{}_I\tau_n^I{}_J)\label{dhs12}
\ee
In the first line, one can use the reality condition (\ref{reality2}) to write $\bar\ep^A\Gamma^{\lambda\eta}B^{-1}\psi^{ J\dag}\phi_{AI}$ as $-\bar\psi^J\Gamma^{\lambda\eta}\ep_A\phi^A_I$; Then,
using the commutator $\tau_m^J{}_I\tau_n^I{}_K-\tau_n^J{}_I\tau_m^I{}_K=f_{mn}{}^p\tau_p^J{}_K$, it is easy to prove that the right hand side of (\ref{dhs12}) vanishes. This finishes the proof of (\ref{dhg}).

We now consider the super-variation on the second equation of (\ref{eqs3}):
\begin{equation}\label{dhphim}
0=\d\bigg(D^2\phi_m-\frac{i}{2}C^\nu(\bar\chi^B_{n}\Gamma_\nu\chi_{Bp})f^{np}{}_m
+\frac{i}{2}C^\nu(\bar\psi^J\Gamma_\nu\psi_I)\tau_m{}^I{}_J+2C^2
\bar\phi^J_A\phi^A_I\phi^n(\tau_{(m}\tau_{n)})^I{}_J\bigg).
\end{equation}
We shall use the same trick for verifying (\ref{dhg}): \emph{Without} coupling to the matter fields $\psi_I$ and $\phi^A_I$, Eq. (\ref{dhphim}) is reduced to
\e
0&=&\bar\d\bigg(D^2\phi_m-\frac{i}{2}C^\nu(\bar\chi^B_{n}\Gamma_\nu\chi_{Bp})f^{np}{}_m
\bigg)\nonumber\\
&=&[\bar\d A_\mu,D^\mu\phi]_m+D_\mu([\bar\d A^\mu,\phi]_m)+D^2(\bar\d\phi_m)
-iC_\mu\overline{\bar\d\chi^A_n}\Gamma^\mu\chi_{Ap}f^{np}{}_m\nonumber\\
&=&2i\bar\ep^A\Gamma_{\mu\nu}C^\nu[\chi_A,D^\mu\phi]_m
+i\bar\ep^A\Gamma_{\mu\nu}C^\nu[D^\mu\chi_A,\phi]_m-i\bar\ep^A D^2\chi_{Am}\nonumber\\
&&-iC_\mu(\ep^{A\dag}\Gamma^{\nu\dag}\Gamma^0D_\nu\phi_n
+\frac{1}{12}\ep^{A\dag}\Gamma^\dag_{\nu\rho\s}\Gamma^0H^{\nu\rho\s}_n)\Gamma^\mu\chi_{Ap}f^{np}{}_m
\label{dhphim2}
\ee
In the second line of the above equation, ``$\bar\d$" refers to the supersymmetry transformation (\ref{susy3}). Using the constraint equations $C^\nu D_\nu\phi_n=0$ and $F_{\mu\nu}=H_{\mu\nu\rho}C^\rho$ (see (\ref{eqs2})), one can simplify (\ref{dhphim2}) to give
\begin{equation}\label{dhphim3}
i\bar\ep^A\Gamma_{\mu\nu}[\chi_A,D^\mu\phi]_mC^\nu+i\bar\ep^A\Gamma_{\mu\nu}[D^\mu\chi_A,\phi]_mC^\nu
-i\bar\ep^AD^2\chi_{Am}-\frac{i}{2}\bar\ep^A\Gamma^{\mu\nu}[F_{\mu\nu},\chi_A]_m
\end{equation}
To prove that (\ref{dhphim3}) vanishes, let us look at the EOM of $\chi_{Am}$ (see (\ref{eqs2})),
\e
0&=&\Gamma^\nu D_\nu\chi_{Am}+\Gamma^\nu C_\nu[\phi,\chi_{A}]_m.\label{chi}
\ee
Multiplying the above equation by $\Gamma^\mu D_\mu$, a short calculation gives
\e
0&=&D^2\chi_{Am}+\frac{1}{2}\Gamma^{\mu\nu}[F_{\mu\nu},\chi_A]_m
-\Gamma_{\mu\nu}[\chi_A,D^\mu\phi]_mC^\nu-\Gamma_{\mu\nu}[D^\mu\chi_A,\phi]_mC^\nu.\label{chi3}
\ee
Multiplying  the above equation by $-i\bar\ep^A$, the right-hand side turns out to be exactly the same as (\ref{dhphim3}), so (\ref{dhphim3}) must vanish. This finishes the proof of (\ref{dhphim2}).

After coupling with the scalar multiplets,  Eq. (\ref{chi}) becomes (see also the fifth equation of (\ref{eqs3})):
\e
0&=&\Gamma^\mu D_\mu\chi_{Am}+\Gamma^\mu C_\mu[\phi, \chi_A]_m+\Gamma^\mu C_\mu B^{-1}\psi^{J}\phi_{AI}\tau_{m}{}^I{}_J
-\Gamma^\mu C_\mu\psi_{J}\bar\phi^{I}_A\tau_{m}{}^J{}_I,
\ee
In exactly the same way for deriving (\ref{chi3}), we can show that
\e
0&=&D^2\chi_{Am}+\frac{1}{2}\Gamma^{\mu\nu}[F_{\mu\nu},\chi_A]_m
-\Gamma_{\mu\nu}[\chi_A,D^\mu\phi]_mC^\nu-\Gamma_{\mu\nu}[D^\mu\chi_A,\phi]_mC^\nu\nonumber\\
&&+\Gamma^{\mu\nu}C_\nu[(B^{-1}D_\mu \psi^{J\dag})\phi_{AI}+B^{-1}\psi^{J\dag}D_\mu\phi_{AI}
-D_\mu\psi_I\bar\phi^J_A-\psi_ID_\mu\bar\phi^J_A]\tau_m^I{}_J.\label{chi4}
\ee

We now begin to calculate (\ref{dhphim}); It can be written as
\e\label{dhphim4}
0&=&[\d A_\mu,D^\mu\phi]_m+D_\mu([\d A^\mu,\phi]_m)+D^2(\d\phi_m)
-iC_\mu\overline{\d\chi^A_n}\Gamma^\mu\chi_{Ap}f^{np}{}_m\nonumber\\
&&+\frac{i}{2}C^\nu[(\bar\psi^J\Gamma_\nu\d\psi_I)+(\overline{\d\psi^J}\Gamma_\nu\psi_I)]\tau_m{}^I{}_J\nonumber\\
&&+2C^2
[(\d\bar\phi^J_A)\phi^A_I\phi^n+\bar\phi^J_A(\d\phi^A_I)\phi^n+\bar\phi^J_A\phi^A_I(\d\phi^n)](\tau_{(m}\tau_{n)})^I{}_J
\ee
Using the relation between the supersymmetry transformations (\ref{susy3}) and (\ref{susy5}), we can write (\ref{dhphim4}) as
\e\label{dhphim5}
0&=&[\bar\d A_\mu,D^\mu\phi]_m+D_\mu([\bar\d A^\mu,\phi]_m)+D^2(\bar\d\phi_m)\nonumber\\
&&-iC_\mu\overline{\bar\d\chi^A_n}\Gamma^\mu\chi_{Ap}f^{np}{}_m-iC_\mu\overline{\d'\chi^A_n}\Gamma^\mu\chi_{Ap}f^{np}{}_m
\nonumber\\
&&+\frac{i}{2}C^\nu[(\bar\psi^J\Gamma_\nu\d\psi_I)+(\overline{\d\psi^J}\Gamma_\nu\psi_I)]\tau_m{}^I{}_J\nonumber\\
&&+2C^2
[(\d\bar\phi^J_A)\phi^A_I\phi^n+\bar\phi^J_A(\d\phi^A_I)\phi^n+\bar\phi^J_A\phi^A_I(\d\phi^n)](\tau_{(m}\tau_{n)})^I{}_J,
\ee
where ``$\bar\d$" and ``$\d$" refer to the super-variations in (\ref{susy3}) and (\ref{susy5}), respectively, and in the second line
\e
\d'\chi_{An}&=&\Gamma_\lambda\ep_BC^\lambda(\bar\phi^J_A\phi^B_I+\bar\phi^{BJ}\phi_{AI})\tau_n{}^I{}_J.
\ee
Using the results for proving (\ref{dhphim2}), and using Eq. (\ref{chi4}), (\ref{dhphim5}) can be converted into
\e\label{dhphim6}
0&=&i\bar\ep^A\Gamma^{\mu\nu}C_\nu[(B^{-1}D_\mu\psi^{J\dag})\phi_{AI}+B^{-1}\psi^{J\dag}D_\mu\phi_{AI}
-(D_\mu\psi_I)\bar\phi^J_A-\psi_ID_\mu\bar\phi^J_A]\tau_m^I{}_J\nonumber\\
&&-iC_\mu\bar\chi^A_n\Gamma^\mu\Gamma_\lambda\ep_BC^\lambda(\bar\phi^J_A\phi^B_I+\bar\phi^{BJ}\phi_{AI})\tau_p{}^I{}_Jf^{np}{}_m\nonumber\\
&&+\frac{i}{2}C_\mu[\bar\psi^J\Gamma^\mu(-2\Gamma^\nu\ep_A D_\nu \phi^A_I-2\Gamma_\lambda\ep_AC^\lambda\tau^{mJ}{}_I\phi_m\phi^A_J)]\tau_m{}^I{}_J\nonumber\\
&&+\frac{i}{2}C_\mu[(-2\Gamma^\nu\ep_A D_\nu \phi^A_I-2\Gamma_\lambda\ep_AC^\lambda\tau^{mJ}{}_I\phi_m\phi^A_J)^\dag\Gamma^0\Gamma^\mu\psi_I]\tau_m{}^I{}_J\nonumber\\
&&+2C^2
[(i\bar\psi^J\ep_A)\phi^A_I\phi^n+\bar\phi^J_A(i\bar\ep^A\psi_I)\phi^n+\bar\phi^J_A\phi^A_I(-i\bar\ep^B\chi_{B}^n)](\tau_{(m}\tau_{n)})^I{}_J
\ee
To prove above equation, we consider the following EOM (see the fourth equation of (\ref{eqs3})):
\e
0&=&\Gamma^\mu D_\mu\psi_I+\Gamma^\mu C_\mu\tau^{mJ}{}_I\phi_m\psi_J-2\Gamma^\mu C_\mu\chi_{Am}\tau^{mJ}{}_I\phi^A_J
\ee
Multiplying the above equation by $C_\nu\Gamma^\nu$, and using $C^\mu D_\mu\psi_I=0$ (see the last line of (\ref{eqs3})), we obtain
\e
0&=&-C_\nu\Gamma^{\mu\nu} D_\mu\psi_I+ C^2\tau^{mJ}{}_I\phi_m\psi_J-2 C^2\chi_{Am}\tau^{mJ}{}_I\phi^A_J\label{dhphim7}
\ee
Substituting (\ref{dhphim7}) into the first line of (\ref{dhphim6}), a straightforward calculation shows that the right-hand side of (\ref{dhphim6}) vanishes. This finishes the calculation of (\ref{dhphim}).

We now try to calculate the super-variation of the first equation of (\ref{eqs3}):
\e
0&=&\d[D^2\phi^A_I-iC^\nu(\bar\chi^{A}_m\Gamma_\nu\psi_J)\tau^{mJ}{}_I
+C^2\phi^A_K\phi_m\phi_n\tau^{mK}{}_J\tau^{nJ}{}_I\nonumber\\
&&+C^2(\bar\phi^K_B\phi^A_L+\bar\phi^{AK}\phi_{BL})\phi^B_J\tau_{m}{}^L{}_K\tau^{mJ}{}_I]\nonumber\\
&=&\d A^n_\mu(-\tau_n^J{}_I)D^\mu\phi^A_J+D^\mu[\d A^n_\mu(-\tau_n^J{}_I)\phi^A_J]+D^2(\d\phi^A_I)\nonumber\\
&&-iC^\nu[(\overline{\d\chi^{A}_m}\Gamma_\nu\psi_J)+(\bar\chi^{A}_m\Gamma_\nu\d\psi_J)]\tau^{mJ}{}_I
\nonumber\\
&&+C^2[(\d\phi^A_K)\phi_m\phi_n+2\phi^A_K(\d\phi_{(m})\phi_{n)}]\tau^{mK}{}_J\tau^{nJ}{}_I
\nonumber\\
&&+C^2(\bar\phi^K_B\phi^A_L+\bar\phi^{AK}\phi_{BL})(\d\phi^B_J)\tau_{m}{}^L{}_K\tau^{mJ}{}_I\nonumber\\
&&+C^2[(\d\bar\phi^K_B)\phi^A_L+\bar\phi^K_B(\d\phi^A_L)
+(\d\bar\phi^{AK})\phi_{BL}+\bar\phi^{AK}(\d\phi_{BL})]\phi^B_J\tau_{m}{}^L{}_K\tau^{mJ}{}_I\label{dphia1}
\ee
Substituting the supersymmetry transformations (\ref{susy5}) into (\ref{dphia1}), and after some work, (\ref{dphia1}) reads
\e
&&-i(\bar\ep^B\Gamma_{\mu\nu}D^\mu\chi^n_B)C^\nu\phi^A_J\tau_m^J{}_I\nonumber\\
&&+i\bar\ep^AD^2\psi_I\nonumber\\
&&+i(\bar\ep^A\Gamma^{\mu\nu}\psi_J)C_\nu D_\mu\phi^m\tau_m^J{}_I\nonumber\\
&&-\frac{i}{2}(\bar\ep^A\Gamma^{\mu\nu}\psi_J)F^m_{\mu\nu}\tau_m^J{}_I\nonumber\\
&&-2i(\bar\ep^B\chi_{B}^{(n})\phi^{m)}C^2\phi^A_K\tau_{n}^{K}{}_J\tau_{m}^{J}{}_I
+2i(\bar\chi^{Am}\ep_B)C^2\phi^n\phi^B_K(\tau_n\tau_m)^K{}_I\nonumber\\
&&+i(\bar\ep^A\psi_K)C^2\phi^n\phi^m\tau_n^K{}_J\tau_m^J{}_I\nonumber\\
&&+iC^2[(\bar\psi^K\ep_B)\phi^A_L\phi^B_J+(\bar\ep^A\psi_L)\bar\phi^K_B\phi^B_J
+(\bar\psi^K\ep^A)\phi_{BL}\phi^B_J+(\bar\ep_B\psi_L)\bar\phi^{AK}\phi^B_J]\tau_m^L{}_K\tau^{mJ}{}_I\nonumber\\
&&-2i(\bar\ep^B\Gamma_{\mu\nu}\chi^n_B)C^\nu D^\mu\phi^A_J\tau_n^J{}_I
+2i(\bar\chi^{Am}\Gamma_{\mu\nu}\ep_B)C^\mu D^\nu\phi^B_J\tau_m^J{}_I\label{dphia2}
\ee

The first line of (\ref{dphia2}) is related to the EOM of $\chi^n_B$. Multiplying the EOM of $\chi^n_B$ (see the fifth equation of (\ref{eqs3})) by $(\phi^A_J\tau_n^J{}_IC^\nu)i\bar\ep^B\Gamma_\nu$, we have
\e
0&=&(\phi^A_J\tau_n^J{}_IC^\nu)i\bar\ep^B\Gamma_\nu\bigg(\Gamma_\mu D^\mu\chi^{n}_B+\Gamma_\mu C^\mu[\phi, \chi_B]^n\nonumber\\
&&+\Gamma_\mu C^\mu B^{-1}\psi^{K}\phi_{BL}\tau_{m}{}^L{}_K
-\Gamma_\mu C^\mu\psi_{L}\bar\phi^{K}_B\tau_{m}{}^L{}_K\bigg)\label{aa}
\ee
Using the reality condition (\ref{reality2}), we obtain $(\bar\ep^B B^{-1}\psi^{K\dag})\phi_{BL}=(\bar\psi^K\ep_B)\phi^B_L$; On the other hand, we have $C_\mu D^\mu\chi^{n}_B=0$ (see the last line of (\ref{eqs3})). Using these two equation, one can convert equation (\ref{aa}) into the form
\e
&&-(i\bar\ep^B\Gamma_{\mu\nu} D^\mu\chi^{n}_B)C^\nu \phi^A_J\tau_n^J{}_I\label{eomchi2}\\
&=&-i\bar\ep^B\chi_{Bm}C^2\phi_p\phi^A_Jf^{pm}{}_n\tau^{nJ}{}_I
-[i(\bar\psi^K\ep_B)\phi^B_L\phi^A_J
-i(\bar\ep^B\psi_L)\bar\phi^K_B\phi^A_J]C^2\tau_n^{L}{}_K\tau^{nJ}{}_I\nonumber
\ee

The second line of (\ref{dphia2}) can be taken care of by using the EOM of $\psi_I$.  Multiplying the EOM of $\psi_I$ (see the fourth equation of (\ref{eqs3})) by $i\bar\ep^A\Gamma^\nu D_\nu$,
\e
0&=&i\bar\ep^A\Gamma^\nu D_\nu\bigg(\Gamma^\mu D_\mu\psi_I+\Gamma^\mu C_\mu\tau^{mJ}{}_I\phi_m\psi_J-2\Gamma^\mu C_\mu\chi_{Bm}\tau^{mJ}{}_I\phi^B_J\bigg).
\ee
Simplifying the above equation gives
\e
i\bar\ep^AD^2\psi_I&=&\bigg(\frac{i}{2}\bar\ep^A\Gamma^{\mu\nu}\psi_JF^m_{\mu\nu}
-i\bar\ep^A\Gamma^{\mu\nu}\psi_JC_\nu D_\mu\phi^m+i(\bar\ep^A\Gamma^{\mu\nu}D_\nu\psi_J)C_\mu\phi^m\nonumber\\
&&-2i(\bar\ep^A\Gamma^{\mu\nu}D_\nu\chi_{B}^m)C_\mu\phi^B_J-2i\bar\ep^A\Gamma^{\mu\nu}\chi_{B}^mC_\mu D_\nu\phi^B_J\bigg)\tau_m^J{}_I.\label{ddpsii}
\ee
One can also take care of the third term of the right-hand side of (\ref{ddpsii}) using the  EOM of $\psi_I$. Multiplying the EOM of $\psi_I$ (see the fourth equation of (\ref{eqs3})) by $i\bar\ep^A\Gamma^\mu C_\mu$,
\e
0&=&i\bar\ep^A\Gamma^\nu C_\nu\bigg(\Gamma^\mu D_\mu\psi_I+\Gamma^\mu C_\mu\tau^{mJ}{}_I\phi_m\psi_J-2\Gamma^\mu C_\mu\chi_{Bm}\tau^{mJ}{}_I\phi^B_J\bigg),
\ee
which can be written as
\e
i(\bar\ep^A\Gamma^{\mu\nu}D_\nu\psi_J)C_\mu\phi^m\tau_m^J{}_I=\bigg(2i\bar\ep^A\chi_{B}^nC^2
\phi^B_K\phi^m-i\bar\ep^A\psi_KC^2\phi^n\phi^m\bigg)(\tau_n\tau_m)^K{}_I.\label{ddpsii2}
\ee
We have used $C_\mu D^\mu\psi_I=0$ (see the last line of (\ref{eqs3})).

Substituting (\ref{eomchi2}) and (\ref{ddpsii}) into (\ref{dphia2}), and using (\ref{ddpsii2}), a straightforward computation shows that (\ref{dphia2}) does vanishes. This complete the calculation of (\ref{dphia1}).

In summary, the super-variation of every EOM vanishes. In other words, Eq. (\ref{danyeom}) is obeyed.

\section{Supercurrents}\label{secsupercurrents}

\subsection{Supercurrent of $\CN=(1,0)$ Theory}
The supercurrent of the $\CN=(1,0)$ theory of Section \ref{secNe10} can be defined as follows,
\e
\bar\ep^Aj_{\mu A}=-ic\bar\chi^A_m\Gamma_\mu\d\chi^m_A+\frac{ic}{2}\overline{\d\psi^I}\Gamma_\mu\psi_I
-\frac{ic}{2}\bar\psi^I\Gamma_\mu\d\psi_I,\label{supercurrent1}
\ee
where $A=1,2$ is an $SU(2)$ R-symmetry index, and $c$ is an overall constant. (In the current of the LP theory \cite{5Lambert2}, $c=-i$.) A short calculation gives
\e
j_{\mu A}&=&-ic[\Gamma_\nu\Gamma_\mu\chi_{Am}D^\nu\phi^m-\frac{1}{2}\Gamma_{\nu\rho}\chi_{Am}H_\mu{}^{\nu\rho m}-\Gamma_\nu\Gamma_\mu\chi_{Bm}C^\nu(\bar\phi^J_A\phi^B_I+\bar\phi^{BJ}\phi_{AI})\tau^{mI}{}_J]\nonumber\\
&&+ic[\Gamma_\nu\Gamma_\mu \psi_I(D^\nu\bar\phi_A^I-C^\nu\phi_m\bar\phi_A^J\tau^{mI}{}_J)]\nonumber\\
&&+ic[\Gamma_\nu\Gamma_\mu B^{-1}\psi^I\ep_{AB}(D^\nu\phi^B_I+C^\nu\phi_m\phi^B_J\tau^{mJ}{}_I)].
\ee
It is straightforward to verify that the current is conserved, using equations of motion and Fierz identities. 

By adding three total derivative terms, one can define the following modified current
\e
\tilde j_{\mu A}=j_{\mu A}+\a_1\Gamma_{\mu\nu}\partial^\nu(\chi_{Am}\phi^m)
+\a_2\Gamma_{\mu\nu}\partial^\nu(B^{-1}\psi^I\phi_{AI})
+\a_3\Gamma_{\mu\nu}\partial^\nu(\psi_I\bar\phi^I_A).
\ee
Here $\a_1$, $\a_2$, and $\a_3$ are constants. The physics remains the same, since $\partial^\mu\tilde j_{\mu A}=\partial^\mu j_{\mu A}=0$ and the total derivative terms do not contribute to the supercharges. If $\tilde j_{\mu A}$ were ``$\Gamma$-traceless", i.e., $\Gamma_\mu \tilde j^\mu_A=0$, it would be possible to define the conserved superconformal current \cite{chen8}
\e
s^\mu_A=\Gamma\cdot x \tilde j^\mu_{A};\label{scft1}
\ee
In fact, one can easily verify that \e\partial_\mu s^\mu_A=\Gamma_\mu \tilde j^\mu_{A}+\Gamma\cdot x \partial_\mu \tilde j^\mu_{A}=0.\ee

A short calculation gives the ``$\Gamma$-trace" of $\tilde j_{\mu A}$,
\e
\Gamma_\mu \tilde j^\mu_A&=&(4ic+5\a_1)\Gamma_\nu\chi_{Am}D^\nu\phi^m-(4ic-5\a_2)\Gamma_\nu B^{-1}\psi^ID^\nu\phi_{AI}-(4ic-5\a_3)\Gamma_\nu \psi_ID^\nu\bar\phi_A^I\nonumber\\
&&-(4ic-10\a_3)\Gamma_\nu\chi_{Bm}C^\nu\bar\phi^J_A\phi^B_I\tau^{mI}{}_J
-(4ic-10\a_2)\Gamma_\nu\chi_{Bm}C^\nu\bar\phi^{BJ}\phi_{AI}\tau^{mI}{}_J\nonumber\\
&&-(4ic-5\a_2+5\a_1)\Gamma_\nu B^{-1}\psi^IC^\nu \phi_m\phi_{AJ}\tau^{mJ}{}_I\nonumber\\
&&+(4ic-5\a_3+5\a_1)\Gamma_\nu \psi_IC^\nu\phi_m\bar\phi_A^J\tau^{mI}{}_J.\label{gammatj1}
\ee
If we set hypermultiplet fields
\e
\psi_I=\phi^I_A=0
\ee
by setting $a_1=0$ in (\ref{red}),  the right-hand side of (\ref{gammatj1}) vanishes, i.e.,
$
\Gamma_\mu \tilde j^\mu_A=0,
$
and it is possible to construct a superconformal current $s^\mu_A$, defined by (\ref{scft1}); as a result, the minimal $(1,0)$ tensor multiplet theory of Section \ref{secmini} may have a superconformal symmetry.

However, if $a_1\neq0$ in (\ref{red}), the right-hand side of (\ref{gammatj1}) fails to vanish without imposing additional constraints on the fields, though one can make either the first line \emph{or} the last three lines to vanish by choosing the values of $\a_1$, $\a_2$, and $\a_3$ properly.
If we set
\e
\a_1=-\a_2=-\a_3=-2ic/5,
\ee
the last three lines of (\ref{gammatj1}) vanish, but the first line remains:
\e
\Gamma_\mu \tilde j^\mu_A&=&2ic\Gamma_\nu\chi_{Am}D^\nu\phi^m-2ic\Gamma_\nu B^{-1}\psi^ID^\nu\phi_{AI}-2ic\Gamma_\nu \psi_ID^\nu\bar\phi_A^I.\label{gammatj3}
\ee
If we set
\e
0=4ic+5\a_1=4ic-5\a_2=4ic-5\a_3,
\ee
the first line of (\ref{gammatj1}) vanishes, and the remaining part is
\e
\Gamma_\mu \tilde j^\mu_A&=&4ic\Gamma_\nu\chi_{Bm}C^\nu\bar\phi^J_A\phi^B_I\tau^{mI}{}_J
+4ic\Gamma_\nu\chi_{Bm}C^\nu\bar\phi^{BJ}\phi_{AI}\tau^{mI}{}_J\nonumber\\
&&+4ic\Gamma_\nu B^{-1}\psi^IC^\nu \phi_m\phi_{AJ}\tau^{mJ}{}_I-4ic\Gamma_\nu \psi_IC^\nu\phi_m\bar\phi_A^J\tau^{mI}{}_J.\label{gammatj2}
\ee
In either case, one cannot construct the conserved superconformal current $s^\mu_A$, meaning that \emph{the general $\CN=(1,0)$ theory does not have a superconformal symmetry}. However, if we impose the additional constraint $\Gamma_\mu \tilde j^\mu_A=0$ or at least $\Gamma_\mu \tilde j^\mu_A|\rm phy\rangle=0$, with $|\rm phy\rangle$ the physical states, it is possible to construct a conserved superconformal $s^\mu_A$, and the $(1,0)$ theory may admit a superconformal symmetry.
(See also (\ref{constaint}) and the discussion below (\ref{constaint}).)

\subsection{Supercurrent of $\CN=(2,0)$ LP Theory}
If the hypermultiplet is also in the adjoint representation of the gauge group (see Section \ref{SecLP}), the supercurrent (\ref{supercurrent1}) becomes
\e
\bar\ep^Aj_{\mu A}=-ic\bar\psi^A_m\Gamma_\mu\d\psi^m_A,
\ee
where $A=1,\ldots,4$ is a $USp(4)=SO(5)$ R-symmetry index, and $\psi^m_A$ is defined by (\ref{usp4psi}), and $\d\psi^m_A$ is defined by the second equation of (\ref{susy8}). (The definitions of all fields of the $(2,0)$ theory can be found in Section \ref{SecLP}.) We see that supercurrent is indeed enhanced from $\CN=(1,0)$ to $(2,0)$. The expression of the $(2,0)$ supercurrent is
\begin{equation}\label{n20crrn1}
j^\mu_{ A}=-ic\bigg(\Gamma^\nu\Gamma^\mu\g^a_A{}^B\psi_B^mD_\nu\phi_m^a-\frac{1}{2}\Gamma_{\nu\rho}\psi_{A}^mH_m^{\mu\nu\rho }-\frac{1}{2}\Gamma^\nu\Gamma^\mu\g^{ab}_A{}^B\psi_{B}^mC_\nu\phi^a_n\phi^b_pf^{np}{}_m\bigg)
\end{equation}
Using the 32-component spinor formalism (see Section \ref{secmini}), it can be written as
\begin{equation}\label{n20crrn2}
j^\mu=ic\bigg(\Gamma^a\Gamma^\nu\Gamma^\mu\psi^mD_\nu\phi_m^a+\frac{1}{2!3!}\Gamma_{\nu\rho\lambda}\Gamma^\mu\psi^mH_m^{\nu\rho\lambda }+\frac{1}{2}\Gamma^\nu\Gamma^\mu\Gamma^{ab}\psi^mC_\nu\phi^a_n\phi^b_pf^{np}{}_m\bigg).
\end{equation}
Here $\mu=0,1,\ldots,5$ and $a=6,\ldots,10$, and $\Gamma^{10}=\Gamma_{0123456789}$.

The three-algebra counterpart of (\ref{n20crrn2}) was constructed in \cite{5Lambert2}; Its expression is
\begin{equation}\label{n20crrn3}
j^\mu_{\rm3alg}=\Gamma^a\Gamma^\nu\Gamma^\mu\psi^mD_\nu\phi_m^a+\frac{1}{2!3!}\Gamma_{\nu\rho\lambda}\Gamma^\mu\psi^mH_m^{\nu\rho\lambda }-\frac{1}{2}\Gamma_\nu\Gamma^\mu\Gamma^{ab}\psi^mC^\nu_o\phi^a_n\phi^b_pf^{onp}{}_m,
\end{equation}
where $f^{onp}{}_m$, being totally antisymmetric in four indices, are the structure constants of three-algebra\footnote{For convenience, we have converted the convention of \cite{5Lambert2} into our convention.}. If we set $c=-i$ in (\ref{n20crrn2}), and make the replacement
\e
C^\nu_of^{onp}{}_m\rightarrow -C^\nu f^{np}{}_m
\ee
in (\ref{n20crrn3}), we see that (\ref{n20crrn3}) is exactly the same as (\ref{n20crrn2}).

We now try to calculate the ``$\Gamma$-trace" of the modified current
\e
\tilde j^\mu_{\rm3alg}=j^\mu_{\rm3alg}+\a\Gamma^{\mu\nu}\partial_\nu(\phi^a_m\Gamma^a\psi^m),
\ee
where $\a$ is a constant. A short computation gives
\e\label{gammatrace}
\Gamma_\mu\tilde j^\mu_{\rm3alg}=(-4+5\a)\Gamma^\nu\Gamma^a\psi^mD_\nu\phi^a_m
+(2-5\a)\Gamma_\nu\Gamma^{ab}\psi^mC^\nu_o\phi^a_n\phi^b_pf^{onp}{}_m
\ee
Again, no matter how we choose the value of $\a$, the right-hand cannot vanish.
So the general $\CN=(2,0)$ LP theory does not have a superconformal symmetry. 

We now consider the possibility of constructing a superconformal current $s^\mu_{\rm3alg}$ by imposing  an additional constraint on the fields. In (\ref{gammatrace}), if we set $\a=4/5$, and assume that the 3-bracket
\e
\psi^m[C^\nu,\phi^a,\phi^b]_m=\psi^mC^\nu_o\phi^a_n\phi^b_pf^{onp}{}_m=0,\label{constaint}
\ee
or at least that the 3-bracket annihilates the physical states, i.e.,$\psi^m[C^\nu,\phi^a,\phi^b]_m|\rm phy\rangle=0$, then we have $\Gamma_\mu\tilde j^\mu_{\rm3alg}=0$ or $\Gamma_\mu\tilde j^\mu_{\rm3alg}|\rm phy\rangle=0$. As a result, it is possible to construct the conserved superconformal current $s^\mu_{\rm3alg}$, and the $\CN=(2,0)$ LP theory may have a superconformal symmetry. It would be interesting to investigate the physical significance of the additional constraint (\ref{constaint}).

\end{document}